\documentclass[reqno,12pt]{amsart}

\usepackage{amsfonts}
\usepackage{bbm}
\usepackage[comma, authoryear]{natbib}
\usepackage{enumerate}
\usepackage{epsfig}
\usepackage{graphicx}
\usepackage{caption}
\usepackage[utf8]{inputenc}
\usepackage{amsmath}
\usepackage{amssymb}
\usepackage{mathrsfs}
\usepackage{float}
\usepackage{algorithm}
\usepackage{algpseudocode}
\usepackage{tabularx,booktabs}
\usepackage{mathtools}
\usepackage{subcaption}
\usepackage{cleveref}
\numberwithin{equation}{section}

\newcolumntype{Y}{>{\centering\arraybackslash}X}

\newcolumntype{b}{Y}
\newcolumntype{s}{>{\hsize=.5\hsize}Y}

\newtheorem{thm}{Theorem}

\newtheorem{remark}{Remark}

\algnewcommand\algorithmicto{\textbf{to}}
\algrenewtext{For}[3]%
{\algorithmicfor\ #1 = #2 \algorithmicto\ #3 \algorithmicdo}

\newfloat{subroutine}{htbp}{loa}
\floatname{subroutine}{Subroutine}
\makeatletter\newcommand\l@subroutine{\@dottedtocline{1}{1.5em}{2.3em}}\makeatother

\newfloat{subroutineb}{htbp}{loa}
\floatname{subroutineb}{Subroutine}
\makeatletter\newcommand\l@subroutineb{\@dottedtocline{1}{1.5em}{2.3em}}\makeatother

\begin{document}
\title[Branching Particle Pricers]{Branching Particle Pricers with Heston Examples}

\author[M. Kouritzin]{Michael A. Kouritzin}
\address{Michael A. Kouritzin\\
	Department of Mathematical and Statistical Sciences\\
	University of Alberta  \\
	Edmonton (Alberta)\\
	Canada T6G 2G1}
\email{michaelk@ualberta.ca
	\newline\indent {\it URL:} http://www.math.ualberta.ca/Kouritzin\_M.html}

\author[A. MacKay]{Anne MacKay}
\address{Anne MacKay\\
	Department of Mathematics\\
	Universit\'{e} du Qu\'{e}bec \`{a} Montr\'{e}al \\
	Montreal (Quebec)\\
	Canada H3C 3P8}
\email{mackay.anne@uqam.ca}

\thanks{Partial funding in support of this work
	was provided by NSERC discovery grants and FRQNT research support for new academics.}
\renewcommand{\subjclassname}{\textup{2010} Mathematics Subject Classification}
\keywords{American Options, Sequential Monte Carlo, Branching Processes,
Heston Model, Stochastic Approximation.}
\dedicatory{University of Alberta and UQAM} \maketitle

\begin{abstract}
The use of sequential Monte Carlo within simulation for path-dependent option pricing is proposed and evaluated. 
Recently, it was shown that explicit solutions and importance sampling are valuable for efficient simulation of spot price and volatility, especially for purposes of path-dependent option pricing. 
The resulting simulation algorithm is an analog to the weighted particle filtering algorithm that might be improved by resampling or branching. 
Indeed, some branching algorithms are shown herein to improve pricing performance substantially while some resampling algorithms are shown to be less suitable in certain cases. 
A historical property is given and explained as the distinguishing feature between the
sequential Monte Carlo algorithms that work on path-dependent option pricing and those that do not.
In particular, it is recommended to use the  so-called effective particle branching algorithm within importance-sampling Monte Carlo methods for path-dependent option pricing.
All recommendations are based upon numeric comparison of option
pricing problems in the Heston model.
\end{abstract}
	

\section{Introduction}\label{intro}

Several years ago, \cite{He} introduced an asset model that
has withstood the test of time and become widely popular.
Its popularity stems from the availability of a closed-form solutions for European option prices and the inclusion of stochastic volatility.
It is, in a way, the ``Black-Scholes model of the stochastic volatility world'' and is widely used to model stock, bond and foreign currency prices.
Recently, \cite{Kouritzin16} introduced explicit weak solutions to the two-dimensional stochastic differential equation
(SDE) solutions of the Heston model that are easy to simulate.

Let $(\Omega,\mathcal F,\{\mathcal F_t\}_{t=0}^T,P)$ be a complete
filtered (risk-neutral) probability space 
supporting (scalar) independent standard Brownian motions $B$, $\beta$. Then, \cite{He} modeled an asset price $S$ and its volatility $V^\frac12$ by
\begin{equation}\label{Heston}
d\left(\begin{array}{c}S_{t}\\V_{t}\end{array}\right)
=\left(\begin{array}{c}\mu S_{t}\\\nu-\varrho V_{t}\end{array}\right)dt+\left(\begin{array}{cc}\sqrt{1-\rho^2}S_{t}V_{t}^{\frac{1}{2}}&\rho S_{t}V_{t}^{\frac{1}{2}}\\0&\kappa V_{t}^{\frac{1}{2}}\end{array}\right)\left(\begin{array}{c}dB_{t}\\d\beta_{t}\end{array}\right),
\end{equation}
with parameters $\mu\in\mathbb R$, $\rho\in[-1,1]$ and $\nu,\varrho,\kappa>0$.
This model has an explicit solution (for all $t > 0 $) with respect to $P$ when $\nu=\frac{n\kappa^2}4$ for some $n\in\{1,2,3,\ldots\}$.
Otherwise, one can still produce an explicit solution with respect to a new probability $\widehat P$ up until (a stopping time when) the volatility drops too low.

Option prices that do not allow closed-form solutions are often priced by Monte Carlo simulation. 
Using \cite{Kouritzin16}'s explicit solutions, it is possible to simulate multiple Heston paths $\{(S_t^j,V_t^j),\ t\ge 0\}_{j=1}^N$ efficiently. 
For reasons that will become apparent in the next paragraphs, we often refer to each of these simulated paths as \textit{particles}. 
Each particle $(S_t^j,V_t^j)$ has the desired distribution (the one under which we want to price) under the pricing measure $P$, and the associated likelihood (or Radon-Nikodym derivative) $L_t^j=\frac{P}{\widehat P^j}\big|_{\mathcal F_t}$ can be seen as its weight.
For this reason, the simulation algorithm for option pricing of \cite{Kouritzin16} can be thought of as the analog of a weighted particle filter in sequential Monte Carlo methods, producing a weighted empirical measure of the form $ \sigma_{[0,t]}^N=\frac1N\sum_{j=1}^N L_t^j \delta_{(S_{[0,t]}^j,V_{[0,t]}^j)}$ at each time step.
Here, $\delta_{(S_{[0,t]}^j,V_{[0,t]}^j)}$ denotes Dirac measure and
$(S_{[0,t]}^j,V_{[0,t]}^j)$ denotes the path of the $j^{th}$-particle's price and volatility
over $[0,t]$ but held constant from $t$ on.
The purpose of the present paper is to discuss the use of resampling and branching in the option pricing procedure in much the same manner as resampling and branching are used in sequential Monte Carlo methods.

The weighted particle filter (credited to \citet{Hand}, 
\citet{HaMa}) also produces an object of the form of $\sigma_{[0,t]}^N$ as an
approximation of the unnormalized filter in problems like tracking and model selection.
While the application and the exact likelihood form differ,
one can think of the overall object in the same way in the context of Monte Carlo Heston simulation.
The particle likelihoods $\{L_t^j\}_{j=1}^N$ provide relative odds that the simulated 
paths $\{(S_{[0,t]}^j,V_{[0,t]}^j)\}_{j=1}^N$ are good representation of the paths $(S_{[0,t]},V_{[0,t]})$ of the underlying model.
However, there is a well known problem in particle filtering (or sequential Monte Carlo):
If left alone, as $t$ increases, many of the simulated particles $\{(S_t^i,V_t^i)\}_{i=1}^N$ move away from the underlying model $(S_t,V_t)$, effectively shrinking the particle system to fewer and fewer relevant particles (i.e. particles that are good representations of the model).

The solution in sequential Monte Carlo was to resample or branch  the particles regularly in an unbiased manner so as to kill bad particles while replicating the good ones with high probability.
The first and still most popular algorithm, called the \emph{bootstrap algorithm},
introduced by \citet{GoSaSm}, simply redistributes the different particles independently back to their last observed locations according to their relative likelihoods.
This algorithm was later improved by the
residual resampling of \citet{LiuChen98},
stratified resampling of \citet{Kita96},
combined residual-stratified resampling discussed in \citet{DoCaMo}
and systematic resampling of \cite{CaClFe}.
In all cases, enough resampling was done that the weights $L_t^i$ could be
reset to $1$ after resampling.
A different approach was taken in \cite{BranchDraft}.
Here, particles are branched individually rather than resampled collectively and 
branching only occurs when there is sufficient need.
Indeed, the cost of rebalancing the system by resampling and branching is the
introduction of noise that immediately degrades estimates. Too much resampling
or branching is thus undesirable.
With this limited branching, weights are kept after branching.
Different branching methods, the residual, combined, dynamic and effective particle
branching algorithms, were presented and shown to outperform all of the resampled
particle algorithms mentioned above on tracking and model selection (see \cite{BranchDraft}).
The simplest of these algorithms, the residual algorithm, has also been analyzed in
\cite{BranchRates}.

Herein, we focus on two of these new branching algorithms and introduce them in the Heston simulation framework of \cite{Kouritzin16}. 
Separately, we show empirically that the bootstrap algorithm is less suitable for pricing options that are strongly path-dependent, mainly because its excessive resampling affects the distribution of path estimates.
Conversely, we demonstrate easy deployment and effectiveness of the new branching algorithms
and recommend the use of the so-called effective particle branching within simulation-based,
path-dependent option pricing problems when $\nu\ne \frac{n\kappa^2}4$ for
any $n\in\mathbb N$ (for there is no need to consider branching or resampling when $\nu= \frac{n\kappa^2}4$).
To do so, we focus on pricing path-dependent options and options allowing for early exercise.

Path-dependent options, including American, Asian and swap options, can be effectively priced using simulation and dynamic programming (DP).
Paths of the underlying option prices and other relevant market variables (e.g. stochastic volatility) are first simulated up to maturity of the option.
Then, the DP algorithms run backwards in time through each simulation (i.e.\ particle) comparing the payoff resulting from immediate exercise to
the discounted future (risk-neutral) expected payout if the option is not exercised.
In particular, American (and other continuously-executable) options are approximated discretely, resulting in Bermudan-style options whose optimal exercise time is determined recursively starting at the option maturity date.
A key breakthrough in this approach by \cite{LoSc} and others (see \cite{Carriere}, \cite{TsitVanRoy}) was the realization that
this conditional expectation of discounted future payout could be estimated
using the simulated particle paths and, further, that the (uncomputable) 
conditional expectations could be approximated by easily-computable projections.
Initially, least-squares regression was used in this projection process.
However, this requires solving an often-ill-conditioned system of linear equations.
\cite{Kouritzin16} then proposed replacing regression with one-step stochastic approximation (SA), which can allow more efficient, accurate pricing without numerical stability issues, but whose performance is sensitive to the chosen step gain.
To reduce the sensitivity of the algorithm to the selection of the step gain, we recommend a modification of the algorithm based on \cite{polyak1992acceleration}.
Herein, we further develop this Monte Carlo pricing approach, started by \cite{LoSc}, by introducing importance sampling into the approach.
Finally, we demonstrate the advantage of resampling through the use of branching algorithms over resampling with interacting algorithms such as the multinomial bootstrap.

The rest of this paper is laid out as follows:
In the next section, we recall the Heston model's weak solutions
and present the weighted Heston simulation algorithm in a framework suggestive of
resampling or branching.
In Section 3, we recall the bootstrap algorithm and explain why it is not suitable for American-style option pricing.
We also introduce the branching algorithms and insert them into the weighted Heston
algorithm to  yield new \emph{combined} and \emph{effective particle branching} Heston simulators.
In Section 4, we insert these branching simulators into the SA/DP option pricing algorithm.
Section 5 consists of our empirical results regarding option pricing.
Finally, Section 6 contains our conclusions.


\section{Explicit and Weighted Heston Algorithms}

\cite{He} developed a closed form solution for European call options in a specific stochastic volatility model.
\cite{BroadieKaya:2006} then came up with an exact simulation method
for the Heston model.
Both of these methods are based upon characterizing and evolving the distribution
of $(S_t,V_t)$.
Later, \cite{Kouritzin16} showed that the Heston SDEs were actually explicitly
solvable in a way that is very convenient for simulation, which facilitates
the pricing of exotic options whose prices cannot be written in closed-form.
Specifically, if we let
\begin{equation}
\nu_\kappa=\frac{n\kappa^2}4,\ 
\mu_\kappa=\mu+\frac{\rho}{\kappa}\left(\nu_\kappa-\nu\right),
\label{numuk}
\end{equation}
for some $n\in\mathbb N$, then the following result is developed in \cite{Kouritzin16}:
\begin{thm} \label{Theorem2}
	Let $\varepsilon\in(0,1)$; $T>0$; $V_0, S_0$ be given random variables with $V_0>\varepsilon$; and
	$\{W^1,\ldots,W^n,B\}$ be independent standard Brownian motions on $(\Omega,\mathcal F,\{\mathcal F_t\}_{t\in [0,T]},P)$. Let also
	\begin{align}
	\widetilde S_t &= S_0
	\exp\bigg(\sqrt{1-\rho^2}\int_0^t \widetilde V_s^\frac12 dB_s
	+\left[\mu-\frac{\nu\rho}\kappa\right] t \nonumber\\
	&\qquad +\left[\frac{\rho\varrho}\kappa-\frac12\right]
	\int_0^t \widetilde V_s ds+\frac\rho\kappa (\widetilde V_t-V_0)
	\bigg)\label{ExplicitStful}\\
	\widetilde V_t&= \sum_{i=1}^n(Y_t^i)^2,
	~~ \eta_\varepsilon=\inf\left\{t:\widetilde V_t\le\varepsilon\right\}\ \text{ and}\label{ExplicitVtful}\\
	\widetilde L_t&=\exp\left\{ \frac{\nu-\nu_\kappa}{\kappa^2} \left[\ln\left(\frac{\widetilde V_t}{V_0}\right)
	+\int_0^t\frac{\kappa^2-\nu_\kappa-\nu}{2\widetilde V_s}+\varrho\, ds
	\right] \right\}
	\label{L2way}
	\end{align}
	where $Y_t^i=\frac{\kappa}2\int_0^t e^{-\frac{\varrho}2 (t-u)}dW^i_u
	+e^{-\frac{\varrho}2t}Y^i_0$ for $i=1,2,...,n$ with $\{Y^i_0\}_{i=1}^n$ satisfying
	$V_0= \sum\limits_{i=1}^n(Y_0^i)^2$.
	Define
	\begin{align*}
	&\beta_t = \sum_{i=1}^n \int_0^t\frac{Y_u^i}{\sqrt{\sum_{j=1}^n(Y_u^j)^2}}dW^i_u+\int_0^{t\wedge\eta_\varepsilon} \frac{\nu-\nu_\kappa}{\kappa \widetilde V_s^\frac12}ds, \qquad\qquad \text{and}\\
	&\widetilde P(A) = E[1_A \widetilde L_{T\wedge\eta_\varepsilon}] \qquad \forall A\in \mathcal F_T.
	\end{align*}
	Then, $\eta_\varepsilon$ is a stopping time and $\widetilde L_{t\wedge\eta_\varepsilon}$ is a likelihood (i.e.\ a non-negative
	$L^r$-martingale with respect to $P$ for any $r>0$ such that $E[\widetilde L_{t\wedge\eta_\varepsilon}]=1$
	for all $t\ge0$).
	Moreover, $(B,\, \beta)$ are independent standard Brownian motions and
	\begin{align}
	d\left(\begin{array}{c}\widetilde S_{t}\\\widetilde V_{t}\end{array}\right)=
	\begin{cases}
	\left(\begin{array}{c}\mu\widetilde S_{t}\\
	\nu-\varrho \widetilde V_{t}\end{array}\right)dt +\left(\begin{array}{cc}\sqrt{1-\rho^2} \widetilde S_{t}\widetilde V_{t}^{\frac{1}{2}}&\rho \widetilde S_{t}\widetilde V_{t}^{\frac{1}{2}}\\
	0&\kappa \widetilde V_{t}^{\frac{1}{2}}\end{array}\right)\left(\begin{array}{c}dB_{t}\\
	d \beta_{t}\end{array}\right),
	& t\le \eta_\varepsilon\\
	\left(\begin{array}{c}\mu_\kappa \widetilde S_{t}\\\nu_\kappa-\varrho \widetilde V_{t}\end{array}\right)dt+\left(\begin{array}{cc}\sqrt{1-\rho^2}\widetilde S_{t}\widetilde V_{t}^{\frac{1}{2}}&\rho \widetilde S_{t}\widetilde V_{t}^{\frac{1}{2}}\\0&\kappa \widetilde V_{t}^{\frac{1}{2}}\end{array}\right)\left(\begin{array}{c}dB_{t}\\d \beta_{t}\end{array}\right),
	& t> \eta_\varepsilon
	\end{cases}
	\label{WeakHestonPwidetilde}
	\end{align}
	on $[0,T]$ with respect to $\widetilde P$.
\end{thm} 
If we take $n=\left\lfloor\frac{4\nu}{\kappa^2}+\frac12\right\rfloor\vee 1$ in (\ref{numuk})
(or equivalently in (\ref{ExplicitVtful})), then
\begin{equation*}\label{SimpHeston}
d\left(\begin{array}{c}\widetilde S_{t}\\\widetilde V_{t}\end{array}\right)
=\left(\begin{array}{c}\mu_\kappa \widetilde S_{t}\\\nu_\kappa-\varrho \widetilde V_{t}\end{array}\right)dt+\left(\begin{array}{cc}\sqrt{1-\rho^2}\widetilde S_{t}\widetilde V_{t}^{\frac{1}{2}}&\rho \widetilde S_{t}\widetilde V_{t}^{\frac{1}{2}}\\0&\kappa \widetilde V_{t}^{\frac{1}{2}}\end{array}\right)\left(\begin{array}{c}dB_{t}\\d \beta_{t}\end{array}\right),
\end{equation*}
is called the \emph{closest explicit} Heston model. Then, Theorem \ref{Theorem2} provides a means to produce weighted particles of the desired Heston model until the volatility drops below $\varepsilon$ and to resort to the closest explicit model thereafter.

\begin{remark}[Notation]
	We are using $\widetilde S,\widetilde V$ for solutions to the closest explicit Heston model under $P$, reserving $S,V$ for the solution to (\ref{Heston}). 
	Henceforth, we will use $\widetilde \beta_t=\sum\limits_{i=1}^n \int_0^t\frac{Y_u^i}{\sqrt{\sum_{j=1}^n(Y_u^j)^2}}dW^i_u$ and $\beta_t=\widetilde \beta_t+\int_0^{t\wedge\eta_\varepsilon} \frac{\nu-\nu_\kappa}{\kappa \widetilde V_s^\frac12}ds$.
\end{remark}

It is necessary to stop (at $\eta_\varepsilon$) before the volatility gets too small.  
Indeed, when the volatility  reaches $V_t^\frac12=0$, the closest explicit and general Heston volatility equations become deterministic
\begin{equation*}
d\widetilde V_t=\nu_\kappa dt,\ \ \ \ dV_t=\nu dt
\end{equation*}
and it is obvious which solution one has.
The model distributions are then singular to each other when $\nu_k\ne \nu$.

Suppose $\{(\widetilde L^j,\widetilde S^j,\widetilde V^j,\eta_\varepsilon^j)\}_{j=1}^N$ are i.i.d.\ copies of $(\widetilde L,\widetilde S,\widetilde V,\eta_\varepsilon)$, where
$\widetilde S,\widetilde V,\widetilde L,\eta_\varepsilon$ are as in Theorem \ref{Theorem2}, and $\widetilde S_{[0,T]}^j,\widetilde V_{[0,T]}^j$ denotes the 
(continuous) paths of the $j^{th}$ particle.
Then, by the (measure-valued) strong law of large numbers, we find that
\begin{equation}
\lim_{N\rightarrow\infty}\frac1N\sum_{j=1}^N\widetilde L^j_{T\wedge\eta_\varepsilon^j}
\delta_{\widetilde S_{[0,T]}^j,\widetilde V_{[0,T]}^j}= \mathcal L(S_{[0,T]},V_{[0,T]})
\text{ a.s.,} 
\label{eq:SLLN}
\end{equation}
where the right hand side means the distribution of $(S_{[0,T]},V_{[0,T]})$
from (\ref{WeakHestonPwidetilde}). 
However, the weights $\{\widetilde L^j_{T\wedge\eta_\varepsilon}\}_{j=1}^N$ will 
typically be very uneven so the effective number of particles, approximated here by the ratio
$ N_T^{eff} \coloneqq \left(\sum\limits_{k=1}^{ N}\widetilde{L}_T^k\right)^2\Big/\sum\limits_{k=1}^{ N}
\left(\widetilde { L}_T^k\right)^2$,
is much smaller than $N$ and the convergence rate is slow (in $N$). 
We will introduce resampling and branching to prevent this from happening.


\section{Sequential Monte Carlo and Branching Heston Algorithm}\label{sec:simulation}

The results of the previous sections can be applied to the efficient simulation of Heston paths and their associated likelihood, which are then used to weight each path in option price calculations. 
In order to avoid the case in which a small number of paths have significantly higher likelihoods than the rest, thus increasing the variance of the estimate and reducing the performance of the simulation algorithm, we resort to regular resampling of the particles. 

We first recall the bootstrap algorithm, which will later be shown to be less suitable for American option pricing. 
We then present two branching algorithms that can improve the performance of the Heston simulation procedure of \cite{Kouritzin16}.

\subsection{Bootstrap Algorithm}\label{sec:bootstrap}
The main idea behind the bootstrap algorithm is to construct a multinomial distribution using the likelihoods associated with each particle and to sample from this distribution.
For clarity, in Algorithm \ref{algo:bootstrap}, we summarize the bootstrap algorithm as it might be used within the simulation of the Heston model stock prices and volatilities.
\begin{algorithm}[h!]
	\caption{Bootstrap Algorithm}
	\begin{algorithmic}[1]
		\Procedure{Bootstrap}{$S_0$,$V_0$,$N$}
		
		\State $\{(S^j_0,V^j_0,L^j_0)=(S_0,V_0,1)\}_{j=1}^{N}$
		\State $W_{N+1}=1$
		
		\For{$t$}{1}{$T$}
		
		\Statex \textit{Evolve the particles independently}
		\State Use Theorem 1 and $\{(S^j_{t-1},V^j_{t-1},L^j_{t-1})_{j=1}^{N}\}$ to create $\left\{(\widehat S_{t}^{j},\widehat V_{t}^{j},\widehat L_{t}^{j})\right\}_{j=1}^{N}$.\label{line:evolve_independently}
		
		\State $\mathbb{L}_{t}=\sum\limits _{i=1}^N\widehat{L}^{i}_{t}$
		
		\Statex \textit{Normalize weights}
		\For{$j$}{1}{$N$}
		\State $w^j_{t}=\frac{\widehat{L}^j_{t}}{ \mathbb{L}_{t}}$, $p_j = \sum\limits_{i=1}^j w^i_{t}$
		\EndFor
		
		\State $k = N-1$
		
		\Statex \textit{Resample}
		
		\For{$j$}{$N$}{1}
		\State Simulate $[0,1]$-Uniform $U_j$ 
		\State $W_j = U^{\frac1j}_j W_{j+1}$
		\While{$W_j \leq p_k$}
		\State $k = k-1$
		\EndWhile
		\State $({S}^{j}_{t},{V}^{j}_{t},{L}^{j}_{t})\doteq({\widehat S}^{k+1}_{t},{\widehat V}^{k+1}_{t},1)$
		
		\EndFor
		
		\EndFor
		\EndProcedure
	\end{algorithmic}
	\label{algo:bootstrap}	
\end{algorithm}

\begin{remark}
	The step described in line \ref{line:evolve_independently} was purposely vague so as not to detract from the 
	bootstrap algorithm.
	These Heston simulation details appear below and can be inserted in place of this step.
	However, we argue and show experimentally that this bootstrap algorithm does not perform as well as other resampling algorithms for American-type option pricing and we recommend not using it.
\end{remark}
The problem with the bootstrap algorithm in path-dependent option pricing is that
there is too much resampling noise.
Suppose that we kept track of each particle through all times.
When resampling occurs, a ``child particle'' moves to a ``parent'' particle site and one can create historical paths $\{S^j_{[0,T]},V^j_{[0,T]}\}_{j=1}^N$ by appending the child's path to its parent's.
If we just focus on discrete times, then it is shown in \cite{DeKoMi} that for the bootstrap algorithm,
\begin{equation*}
\lim_{N\rightarrow\infty}\frac1N\sum_{j=1}^N
\delta_{(S_{0}^j,V_{0}^j),(S_{1}^j,V_{1}^j),...,(S_{T}^j,V_{T}^j)}= 
\mathcal L(S_{0},V_{0}) \otimes\mathcal L(S_{1},V_{1})
\otimes\cdots\otimes\mathcal L(S_{T},V_{T}),
\end{equation*}
meaning the limit is not the (discrete-time) process distribution of the desired Heston model
but rather product measure of the marginals, i.e.\ independence.
This means that the historical property \eqref{eq:SLLN} does not hold and estimating the conditional expectation of future discounted payoff can not be determined correctly from the cross-sectional data when the bootstrap is used, which has a significant impact on path-dependent option pricing.
If one mistakenly calculated this estimate with bootstrap, he or she would wind up with a regular expectation instead of a conditional one, which is not useful to many option pricing problems.
A good way to correct for this is to switch to the branching algorithms 
in \cite{BranchDraft}.

We note that there is a second (non-bootstrap) resampling algorithm in \cite{DeKoMi}
that does have the required historical property.
However, this second algorithm is slower and arguably more complicated than the branching algorithms
discussed below.

\subsection{Branching Algorithms}\label{BranAlg}
Compared to collective resampling algorithms such as the bootstrap one presented above, individual branching improves speed by simplifying decisions.
The decision to branch (or kill) a particle (that is, the decision to resample the same particle more than once, or not at all) is based upon the weight of that individual particle and the total weight of all particles, and not on the individual values of the other ones.
Our basic branching framework is described in Algorithm \ref{algo:branching}. 

\begin{algorithm}[h!]
	\caption{Branching Algorithm}\label{algo:branching}
	\begin{algorithmic}[1]
		\Procedure{GeneralBranching}{$S_0$,$V_0$,$N$,$T$,$r$}
		
		\For{$t$}{1}{$T$}
		
		\Statex \textit{Evolve particles independently using Theorem \ref{Theorem2}}
		\State Simulate $\left\{(\widehat S_{t}^{j},\widehat V_{t}^{j},\widehat L_{t}^{j})_{j=1}^{N_{t-1}}\right\}$ from $\{(S^j_{t-1},V^j_{t-1},L^j_{t-1})_{j=1}^{N_{t-1}}\}$. \label{line:basic_evolve_independently}
		
		\Statex \textit{Calculate average weight}
		\State ${A}_{t}=\frac1N\sum_{j=1}^{N_{t-1}}{\widehat L}^{j}_{t}$
		\label{line:At}
		
		\Statex \textit{Check which particles to branch}
		\State $l=0$ 
		
		\For{$j$}{1}{$N_{t-1}$}
		
		\If{$\widehat{L}^{j}_{t}\in \left(\frac1{r_t}A_t,r_tA_t\right)$} \label{line:non_branching_start}
		
		\Statex \textit{Move non-branched particles to final vector}
		\State $(S_t^{j-l},V_{t}^{j-l},{L}_{t}^{j-l}) =({\widehat S_t}^j,{\widehat V_t}^j,\widehat {L}_{t}^{j})$ 
		
		\Else
		
		\State $l=l+1$
		\State $(\widehat S_t^{l},\widehat V_{t}^{l},\widehat {L}_{t}^{l}) =({\widehat S_t}^j,{\widehat V_t}^j,\widehat {L}_{t}^{j})$
		
		\EndIf \label{line:non_branching_end}
		\EndFor
		
		\Statex \textit{Branching part of the algorithm}
		\State $N_t = l$ 
		\State Simulate $\{W^{j}_{t}\}_{j=l+1}^{N_{t-1}}$, with $W_{t}^j\sim \left[\frac{j-l-1}{N_{t-1}-l},\frac{j-l}{N_{t-1}-l}\right]$-Uniform \label{line:simulate_unif}
		\State Let $p$ be a random permutation of $\{l+1,l+2,...,N_{t-1}\}$
		\label{line:YF1}
		\State $U^{j}_{t}=W^{p(j)}_{t}$
		\label{line:YF2}
		
		\For{$j$}{$l+1$}{$N_{t-1}$} \label{line:combined_branch_start}
		
		\State $N_{t}^j=
		\left\lfloor \frac{\widehat{L}^{j-l}_{t}}{{A}_{t}}\right\rfloor+1_{\left\{
			U^{j}_{t}\le \left(\frac{\widehat{L}^{j-l}_{t}}{{A}_{t}}-
			\left\lfloor\frac{\widehat{L}^{j-l}_{t}}{{A}_{t}}\right\rfloor\right)\right\}}$
		
		\For{$k$}{1}{$N^j_t$}
		
		\State $(S_{t}^{N_{t}+k},V_{t}^{N_{t}+k},L_t^{N_{t}+k}) =(\widehat S_{t}^{j-l},\widehat V_{t}^{j-l},A_t)$	
		
		\EndFor 
		\label{line:combined_branch_end}
		
		\State $N_{t}=N_{t}+N_{t}^j$
		
		\EndFor
		
		\EndFor

		\EndProcedure
	\end{algorithmic}
\end{algorithm}

The key steps of Algorithm \ref{algo:branching} are described between lines \ref{line:non_branching_start} and \ref{line:combined_branch_end}.
They serve determine the new number of particles $N_{t}$ and weights $\{L^j_{t}\}_{j=1}^{N_t}$ in an unbiased manner.
The main idea of the branching algorithm is the following.
When the prior weight $\widehat {L}^j_{t}$ for particle $j$ is extreme (that is, outside of a certain interval around the average weight), we do (limited) residual-style branching, which helps to preserve the process distribution.
Particles that are branched result in zero or more particles, which are assigned the average weight $A_t$, are added at the same location as the parent.
In other words, we copy (or kill) the path with extreme prior weight and give the copies, if there are any, the current average weight.
When its prior weight $\widehat {L}^j_{t}$ is not extreme, a particle is not branched and gets to keep its prior weight.
The flexibility in this class of algorithms is how we determine ``extreme".

The resampling parameter $r=(r_t; t \geq 0)$ determines the size of the interval around the average weight $A_t$ outside of which particles are considered extreme and are branched. This parameter can be obtained using different methods (see below) and controls the amount of branching:
If $r_t=\infty$, then there is no branching.
If $r_t=1$, then every particle is branched.
Both of these cases are rarely good.
Instead, one looks for a good fixed $r\in(1,\infty)$ (\emph{combined branching}) or even makes $r_t$ depend upon the system of particles (\emph{effective particle branching}).

When it is necessary to record the whole path and not just its final value, we just append a child's path to that of its parent.
It follows that each child at time $t\ge 2$ has a unique parent at time $t-1$ and grandparent $t-2$ but
a particle at time $t$ might have no or several children at time $t+1$.
Then, the historical paths $\{S^j_{[0,T]},V^j_{[0,T]}\}_{j=1}^{N_T}$ (or a discretization
thereof) can be used for option pricing.
Indeed, in Section 5, we price options using 
\begin{equation}\label{HisSLLN}
\lim_{N\rightarrow\infty}\frac1N\sum_{j=1}^N\widehat L^j_{T\wedge\eta_\varepsilon^j}
\delta_{\widehat S_{[0,T]}^j,\widehat V_{[0,T]}^j}= \mathcal L(S_{[0,T]},V_{[0,T]})
\text{ a.s.} 
\end{equation}

The precise mathematical proof will resemble that of the
strong law of large numbers in Theorem 5.1 in \cite{BranchRates} but is left to future work.
More information on this branching algorithm framework can be found in \cite{BranchDraft}.

One key aspect of these branching algorithms discussed in \cite{BranchDraft} is that
the number of particles do not vary wildly. 
Wild particle number variation could affect performance and cause a method to fail.
Notice that the average weight $A_t$ in line \ref{line:At} is normalized by
the initial number of particles (rather than the current), which forces the
expected number of future particles given the current to be the initial $N$.

We present below two branching options. 
The second one is a refinement of the first one, and it should perform better. 
We mention that the performance of the branching algorithms is tied to how well they control their particle numbers, as well as under what circumstances they will branch more, and direct the reader to \cite{BranchDraft} for more details and other branching algorithms.

\subsubsection{Combined Branching}
This is the basic algorithm used in Algorithm \ref{algo:branching}. 
It uses stratified resampling to help control the number of
particles and improve algorithm performance.
The $\{\rho^{j}_{t}\}_{j=1}^{N_{t-1}}$ are negatively correlated, which is desirable for particle control.
The generation of a large number of new particles will be more likely to be compensated after by the generation of a smaller number of new particles, because of the negative correlation between $\{\rho^{j}_{t}\}_{j=1}^{N_{t-1}}$.

In lines \ref{line:non_branching_start} to \ref{line:non_branching_end} of Algorithm \ref{algo:branching}, we handle the non-branched particles while moving the ones
that will be branched to the \emph{front of the line}.
In line \ref{line:simulate_unif}, we create the required number of stratified uniform random numbers $\{U^j_t\}$
and in lines \ref{line:combined_branch_start} to \ref{line:combined_branch_end}, we branch the particles designated for branching.
We have ${\rho^{j}_{t}=1_{\left\{U^{j}_{t}\le \left(\frac{\widehat{L}^{j}_{t}}{{A}_{t}}-
		\left\lfloor\frac{\widehat{L}^{j}_{t}}{{A}_{t}}\right\rfloor\right)\right\}}}$	
so we are actually using a residual technique (see \cite{BranchDraft}).
However, generating negatively correlated $\{\rho^{j}_{t}\}_{j=1}^{N_{t-1}}$, rather than independent one as in the \emph{Residual branching} algorithm of \cite{BranchDraft} reduces the probability of getting mostly large or mostly small
uniform random numbers. This reduces the variation in $\rho^{j}_{t}$ and thus in the number of particles.
\begin{remark}
	In combined branching, we use a suitable fixed ${r_t=r>1}$.
	The better values are model dependent and our choices are given within each example	in Section 5.
	For most problems, the optimal proportion of paths that are branched at each time step is between $0.05$ and $0.65$.
	For a given problem, $r$ can be chosen by minimizing the root-mean-square error (RMSE), or another measure of the difference, between the theoretical and estimated values of quantities that are known theoretically, such as the expectation of $L_T$ or $S_T$.
	This minimization can be performed using a stochastic gradient descent algorithm.
	In many cases, the RMSE will not be very sensitive to $r$.
	When this is the case, one should choose $r$ to manage the trade off between adding branching noise now and re-distributing the
	particles for better future use.
\end{remark}

\subsubsection{Effective Particle Branching}\label{ssec:effective}
In the prior method, a constant $r_t$ was used for simplicity only.
If one is in need of fast, accurate option pricing, then there are better choices.

The uneven weights that arise from direct use of 
Theorem \ref{Theorem2} result in an
\emph{effective} number of particles, $N^{eff}$, that can be estimated.
In our setting, the estimated number of effective and non-effective particles are:
\begin{equation*}
N^{eff}_{t-1}=\frac{N^2 A^2_{t}}{\sum\limits_{j=1}^{ N_{t-1}}
	\left(\widehat { L}_{t}^j\right)^2}=\frac{
	\left(\sum\limits_{j=1}^{ N_{t-1}}\widehat {L}_{t}^j\right)^2}{\sum\limits_{j=1}^{ N_{t-1}}
	\left(\widehat { L}_{t}^j\right)^2},\ \  N^{noneff}_{t-1}= N_{t-1}- N^{eff}_{t-1}.
\end{equation*}
From the second equality, we see that $N^{eff}_{t-1}=N_{t-1}$ if all $\widehat {L}_{t}^j$
are the same so all particles are equally effective and $N^{eff}_{t-1}=1$ if all but
one of the $\widehat {L}_{t}^j$ were $0$ so there is only one effective particle.
(None of the $\widehat {L}_{t}^j$ can be zero but they can be arbitrarily close.)
Otherwise, it gives us a number somewhere in between that can be interpreted
as the effective number of particles.
It is very reasonable to anticipate better results when branching either
more or fewer particles in the situation there are few effective ones.
A first intuition might lead us to the conclusion that it is better to
branch more in order to obtain more effective particles immediately.
However, if those few particles with high weights happen to be wrong, then
we will likely be creating a large number of copies of these ``bad'' paths. 
We do not assume either a priori but rather, in the \emph{effective} particle branching algorithm, we set
\begin{eqnarray}\label{rceff}
r_t=\frac{ c^{eff} N^{eff}_{t-1}+c^{noneff} N^{noneff}_{t-1}}{ N_{t-1}}
=c^{noneff}+(c^{eff}-c^{noneff})\frac{ N^{eff}_{t-1}}{ N_{t-1}}
\end{eqnarray}
for experimentally determined constants $c^{eff}$, $c^{noneff}>0$
and let the data decide.

\subsection{Weighted and Explicit Heston Simulation}\label{HestSimAlg}

We now turn our attention to line \ref{line:basic_evolve_independently} in the basic branching algorithm
(this step also appears in line \ref{line:evolve_independently} of the bootstrap algorithm), in which $(S^j_{t-1},V^j_{t-1},L^j_{t-1})_{j=1}^{N_{t-1}}$ are evolved to obtain $(\widehat S^j_t,\widehat V^j_t,\widehat L^j_t)_{j=1}^{N_{t-1}}$. Here, we recall how to perform this step using Theorem \ref{Theorem2} (see also \cite{Kouritzin16}). 

Defining constants 
\begin{equation*}
a=\sqrt{1-\rho^2},\ b=\mu-\frac{\nu\rho}{\kappa},\ c=\frac{\rho\varrho}{\kappa}-\frac{1}{2},\
d= \frac{\rho}{\kappa},\ e=\frac{\nu-\nu_\kappa}{\kappa^2},\ f=e\frac{\kappa^2-\nu-\nu_\kappa}{2}, 
\end{equation*}
we find that (\ref{ExplicitStful},\ref{L2way}) can be rewritten as
\begin{align}\label{Stfulabcd}
\widetilde S_t&= \widetilde S_{t-1}
\exp\left\{a\int_{t-1}^t \widetilde V_s^\frac12 dB_s
+b +c \int_{t-1}^t \widetilde V_s~ ds+d\ (\widetilde V_t-\widetilde V_{t-1})\right\}
\\
\label{Lef}
\widetilde L_t&=\widetilde L_{t-1}\exp\left\{ e \left(\ln\left(\frac{\widetilde V_t}{\widetilde V_{t-1}}\right)+\varrho\right) 
+f\int_{t-1}^t\frac{1}{\widetilde V_s}\,ds
\right\},
\end{align}
which simplifies the simulation of the values at a future time step. 
Indeed, the stochastic integral in (\ref{Stfulabcd}) is conditionally (given $\widetilde V$) Gaussian since $\widetilde V$ 
and $B$ are independent so it can be simulated as a centered normal random variable with
variance $a^2\int_{t-1}^t \!\widetilde V_s ds$.
Even the likelihood (\ref{Lef}) avoids stochastic integrals.

There are a number of ways to compute the two deterministic integrals in \eqref{Stfulabcd} and \eqref{Lef}.
\cite{Kouritzin16} mentions that Simpson's $\frac13$ rule with $M=6$ is a good choice. 
In general, the choice of $M$ should depend on the Heston's model parameters, more specifically on the ratio $\frac{\nu}{\kappa^2}$, and on the size of the time steps.
Indeed, if $\frac{\nu}{\kappa^2}$ is low (or in our case, if $\frac{4\nu}{\kappa^2}$ is close to 2) a larger $M$, i.e. a finer discretization, should be used in order for the simulated $V^j_t$'s not to drop too close to zero.
When $\frac{\nu}{\kappa^2}$ is higher, numerical tests have shown that the precision of the estimate is not affected by the use of a smaller $M$ (see Section \ref{ssec:european_numerical_results}).
When this is the case, it is better to use a lower value for $M$ to speed up the algorithm.
Of course, if the time steps used in the simulation are longer, a larger $M$ should be selected to increase the precision of the deterministic integral.
However, since one of the advantages of Algorithm \ref{algo:Weighted} is its speed, we recommend using it to price path-dependent options when the financial market needs to be simulated at short time intervals.

For the presentation of the algorithm, it will be notationally convenient to remove the tildes and to define two more constants
\begin{equation}
\sigma_M=\kappa\sqrt{\frac{1-e^{-\frac\varrho{M}}}{4\varrho}},\text{ and } \alpha_M=e^{-\frac\varrho{2M}}. 
\end{equation}
The resulting procedure to simulate paths of the Heston model using $T$ time steps of length 1 is given in Algorithm \ref{algo:Weighted}.

\begin{remark}
	Algorithm \ref{algo:Weighted} is presented in its most basic form. Its performance can be further improved through the use of antithetic variates for the $Z^{j,i}$'s. Another way to further speed up the algorithm could be to generate a large sample of Normal random variables and to re-use them in such a way to minimize the dependence thus created. The application of this method is left for future work.
\end{remark}

\begin{algorithm}[h!]
	\caption{Weighted simulation}\label{algo:Weighted}
	\begin{algorithmic}[1]
		\Procedure{WeightedSimulation}{$S_0$, $V_0$, $n$, $N$, $M$, $T$}
		
		\State $\left\{(S_0^j,L_0^j,\eta^j_\varepsilon)=(S_0,1,T)\right\}_{j=1}^N$, $\left\{Y_{0}^{j,i}=\sqrt{\frac{V_{0}}{n}}\right\}_{j,i=1}^{N,n}$, $\left\{V_{\frac{k}M}^{j}=0\right\}_{j,k=1}^{N,MT}$ 
		
		\For{$t$}{1}{$T$}

		\State $\{V^j_t = 0\}_{j=0}^N$
		
		\For{$k$}{$M-1$}{0}\label{line:begin}
		
		\State Generate $(0,1)$-Normal random variables $\{Z^{j,i}\}_{j,i=1}^{N,n}$ \label{line:Zi}
		
		\State $\{Y^{j,i}_{t-\frac kM}=\alpha_M Y^{j,i}_{t-\frac{k+1}M} + \sigma_M Z^{j,i}\}_{j,i=1}^{N,n}$
		
		\State $\{V^j_{t-\frac kM} = {V}^j_{t-\frac kM} + \sum_{i=1}^n({Y}^{j,i}_{t-\frac kM})^2\}_{j=1}^N$ 
		
		\EndFor\label{line:end}
		
		\State $\left\{IntV^j=\frac{V^j_{t-1}+4V^j_{t-\frac{M-1}M}+2V^j_{t-\frac{M-2}M}+\cdots+2V^j_{t-\frac{2}M}+4V^j_{t-\frac1M}+V^j_t}{3M}\right\}_{j=1}^N$
		
		\State Generate $N$ $\left(0,a\sqrt{ IntV^j}\right)$-Normal $\{Z^j\}_{j=1}^N$. 
		
		\State $\{S_t^j=S_{t-1}^j\exp(N^j+b+c\,IntV^j +d\ (V^j_t-V^j_{t-1}))\}_{j=1}^N$
		
		\For{$j$}{1}{$N$}		
		
		\If {$t \leq \eta^j_{\varepsilon}}$
		
		\If{ $\min_{k\in\{0,1,...,M-1\}}V^j_{t-\frac{k}M}>\varepsilon$}
		
		\State $Int\frac1{V^j}=\frac{1}{3M}
		\left[\frac{1}{V^j_{t-1}}+\frac{4}{V^j_{t-\frac{M-1}M}}+\frac2{V^j_{t-\frac{M-2}M}}+\cdots+\frac{2}{V^j_{t-\frac{2}M}}+\frac4{V^j_{t-\frac1M}}+\frac1{V^j_t}\right]$
		\State $L_t^j=L_{t-1}^j\exp\left\{ e \left(\ln\left(\frac{V^j_t}{V^j_{t-1}}\right)+\varrho \right)
		+f Int\frac1{V^j}\right\}$\label{line:Lt}
		
		\Else
		\State ${\eta}^j_\varepsilon = t-1$
		\EndIf
		\EndIf	
		\EndFor
		\EndFor
		\EndProcedure
	\end{algorithmic}
\end{algorithm}


\section{Pricing by SA/DP Algorithm} \label{Pricing}

We now turn our attention to pricing American-style options in the Heston model. 
The risk-neutral (pricing) measure we use going forward is the one denoted by $\widetilde P$ in \eqref{WeakHestonPwidetilde}. 
In other words, $\nu \neq \frac{n \kappa^2}{4}$ for any $n \in \mathbb{N}$, and we must resort to Algorithm \ref{algo:Weighted}, along with branching, to simulate paths of $(S,V)$ for pricing purposes.

We consider an option that can be exercised only once, at any time before or at maturity. The payoff process of the option represents the amount received by the investor given that the option is exercised at $t$, for $t \in \{0,1,\ldots,T\}$\footnote{Since we only consider discrete exercise times, we are technically pricing Bermudan options. For ease of notation, we use a subset of $\mathbb{N}$ to denote the possible exercise times, without loss of generality.} We denote this process by $\{Z_t,\ t\ge0\}$ and assume that it satisfies either 
\begin{equation*}
Z_t=p(t,S_t) \qquad\text{or}\qquad Z_t=p(t,R_t)
\end{equation*}
for some non-negative measurable function $p$, where $R_t=\frac1t \sum_{k=1}^t S_t$ is the running average of the underlying asset price. For example, using the notation ${x^+ = \max(x,0)}$, we have that ${p(t,S_t)=e^{-\mu t}(K-S_t)^+}$ for an American put or ${p(t,R_t)=e^{-\mu t}(K-R_t)^+}$ for an Asian put with early exercise. Recall from \eqref{WeakHestonPwidetilde} that $\mu$ denotes the drift of $S_t$ under the risk-neutral measure, and so $e^{-\mu t}$ discounts the payoff from time $t$ to $0$.

\begin{remark}
	In numerical implementations, the running average price $R_t$ should be recursively calculated from the price $S$ by setting $R_0=0$ and 
	\begin{equation*}
	R_t=\frac{t-1}t R_{t-1} +\frac1t S_t
	\end{equation*}
	for $t \in \{1,2,\ldots,T\}$.
\end{remark}

The price at $t=0$ of an American-style option with payoff process $Z_t$, denoted here by $C_0$ is defined as  
\begin{equation}
C_0 = \sup_{\tau\in\mathcal T_{0,T}}\widetilde E[Z_{\tau}],\label{eq:americanprice}
\end{equation}
where $\mathcal T_{t,T}$ denotes the collection of stopping times taking values in $\{t,t+1,...,T\}$ and
$\widetilde E$ denotes expectation with respect to $\widetilde P$. By letting $\tau_t\in\mathcal T_{t,T}$ be such that $\widetilde E[Z_{\tau_t}]=\sup_{\tau\in\mathcal T_{t,T}}\widetilde E[Z_{\tau}]$ 
and $0\le \tau_0\le \tau_1\le\cdots\le \tau_T=T$, we have $C_0 = \widetilde{E}[Z_{\tau_0}]$.\footnote{For more details on pricing American options, see for example Chapter 21 of \cite{bjork2009arbitrage}.}

American option pricing problems therefore involve calculating $\widetilde{E}[Z_{\tau_t}]$ for different values $t \in \{0,\ldots,T\}$, which can be done using Monte Carlo simulations and dynamic programming. This so-called \textit{least-squares Monte Carlo} method has the advantage of being highly flexible (see e.g.\ \cite{LoSc}, \cite{ClPr}). Here, we recall the main ideas of the method and present an algorithm that makes use of stochastic approximation techniques to increase execution speed and numerical stability (see also  \cite{Kouritzin16}). 

As explained below, a key step of the pricing algorithm relies on the projection of the future payoff. \cite{ClPr} used the general assumptions discussed below to ensure that this projection can be done correctly. We first note that the procedure relies on the underlying model being a Markov chain $\{X_t\}_{t=0}^\infty$ and on the payoff being of the form $Z_t = f(t, X_t)$. Therefore, when pricing payoffs depending on the running average, it is necessary to add a third state variable, $R_t$, to our underlying model. We denote by $D = D_S \times D_V \times D_R$ the state space of the Markov process $(S,V,R)$. Then, the following assumptions are required to use projections to approximate $\widetilde{E}[Z_{\tau_t}]$.
\begin{itemize}
	\item
	\textbf{Total:} there are measurable $\mathbb R$-valued functions $(e_k)_{k=1}^\infty$ on $D$  such that $\{e_k(S_t,V_t,R_t)\}_{k=1}^\infty$ is total\footnote{A subset of a Hilbert space is
		total if its span is the entire space.} 
	on $L^2(\sigma(S_t,V_t),1_{\{Z_t>0\}}d\widetilde P)$ for all $t=1,...,T-1$.
	\footnote{\cite{ClPr} also assumed functions $(f_t)_{t=0}^T$ such that 
		$\widetilde E[Z_{\tau_t}|\mathcal F_t]=f_t(S_t,V_t)$ for all $t=0,...,T$.
		However, one can replace $\{\mathcal F_t\}$ with the filtration generated
		by $(S,V)$ and then these functions exist for our payoff process.} 
	\item
	\textbf{Non-singular:} $\widetilde E[e^J(S_t,V_t,R_t)(e^J(S_t,V_t,R_t))'1_{\{Z_t>0\}}]$ is positive definite for all $J\in\mathbb N$, where $e^J=(e_1,...,e_J)'$.
\end{itemize}
The point of these conditions is that we need a suitable collection 
of functions $\{e_k\}_{k=1}^\infty$ that can be used for projection of a future payoff. Typically, $\{e_k(S_t,V_t,R_t)\}_{k=1}^\infty$ is an ordering of the basis functions $\{e_{k_1}(S_t),e_{k_2}(V_t),e_{k_3}(R_t)\}_{k_1, k_2, k_3=1}^\infty$.
Different choices of basis functions are possible. \cite{LoSc} mention the weighted Laguerre, Hermite, Legendre, Chebyshev, Gegenbauer and Jacobi polynomials. \cite{Kouritzin16} also discusses using a modification to the Haar polynomials. Both articles use the weighted Laguerre polynomials in numerical examples, which is what we do in Section 5.

Theoretically, suitable stopping times can be obtained by working backwards from $T$:
\begin{align}
\begin{cases}
&\tau_T=T, \\
&\tau_t=t1_{\{Z_t\ge \widetilde{E}[Z_{\tau_{t+1}}|\mathcal F_t]\}\cap\{Z_t>0\}}+\tau_{t+1}1_{\{Z_t< \widetilde{E}[Z_{\tau_{t+1}}|\mathcal F_t]\}\cup\{Z_t=0\}}, \qquad T < t. 
\end{cases}
\label{eq:stoppingtimes}
\end{align}
\begin{remark}
	Typically, $\widetilde{E}[Z_{\tau_{t+1}}|\mathcal F_t]>0$ so $\cap\{Z_t>0\}$ and $\cup\{Z_t=0\}$ do not affect the recursion in \eqref{eq:stoppingtimes}.
\end{remark}
In practice, the conditional expectations $\{\widetilde{E}[Z_{\tau_{t+1}}|\mathcal F_t]\}_{t=0}^{T-1}$ and stopping times are not immediately computable.
Therefore, we follow \cite{LoSc} and approximate the conditional expectations by projections 
onto the closed linear span of $e^J(S_t,V_t)$
(see also \cite{ClPr}, \cite{Kouritzin16}), which leads to the following approximate stopping times, which depends on the number $J$ of projection functions used
\begin{align*}
\begin{cases}
&\tau_T^J=T, \\
&\tau_t^J=t1_{\{Z_t\ge P^J_t[Z_{\tau^J_{t+1}}]\}\cap\{Z_t>0\}}+\tau_{t+1}^J1_{\{Z_t< P^J_t[Z_{\tau^{J}_{t+1}}]\}\cup\{Z_t=0\}}, \qquad T < t.
\end{cases}
\end{align*}
The problem is still not solved because this projection,
$P^J_t[Z_{\tau^J_{t+1}}]=\alpha^J_t\cdot e^J(S_t,V_t) $, is computed
using coefficients $\alpha^J_t$ that are not known.
However, \cite{LoSc} showed that these coefficients $\alpha^J_t$ can
be estimated using cross-sections of a Monte Carlo particle system via linear regression. \cite{Kouritzin16} then showed that stochastic approximation can be
a favorable alternative to the original least-squares approach.
Neither of these works contemplated using the method in a branching particle system, which is what we do here. 

In addition, to improve the robustness and the performance of the algorithm, we use the average of the coefficients $\alpha_t^J$ in the projection.
Averaging the coefficients was first proposed by \cite{polyak1992acceleration} and is now a widely used method that can speed up the convergence of the stochastic approximation algorithm.
We recall our branching particle system and apply the stochastic approximation algorithm to our setting next.

Hereafter, suppose $\{(L_t^j,S_t^j,V_t^j), t\ge 0\}_{j=1}^N$ is
as in Section \ref{BranAlg} so (\ref{HisSLLN}) holds, where
$(S,V)$ satisfies (\ref{WeakHestonPwidetilde}) and
$Z^j_t=p(t,S^j_t)$ or $Z^j_t=p(t,R^j_t)$, with $R^j_t=\frac1t \sum_{k=1}^t S^j_t$.
Moreover, define the particle-by-particle stopping times
\begin{equation}\label{tauJjdef}
\begin{cases}
&\tau_T^{J,j}=T  \\
&\tau_t^{J,j}=t1_{\{Z^j_t\ge P^{J}_t[Z^j_{\tau^{J,j}_{t+1}}]\}\cap\{Z^j_t>0\}}+\tau_{t+1}^{J,j}1_{\{Z^j_t< P^{J}_t[Z^j_{\tau^{J,j}_{t+1}}]\}\cup\{Z^j_t=0\}}, \qquad t<T.
\end{cases}	
\end{equation}
and suppose
\begin{align*}
A^N_t=\frac1N \sum_{j=1}^N A_j, &\qquad\text{ where } A_j=
\frac{L_t^je^J(S^j_t,V^j_t)e^J(S^j_t,V^j_t)' 1_{\{Z^j_t>0\}}}{\frac1N \sum_{i=1}^N1_{\{Z^i_t>0\}}},\\ 
b^N_t=\frac1N \sum_{j=1}^N b_j, &\qquad\text{ where } b_j=\frac{L_t^j Z^j_{\tau^{J,j}_{t+1}}e^J(S^j_t,V^j_t) 1_{\{Z^j_t>0\}}}{\frac1N \sum_{i=1}^N1_{\{Z^i_t>0\}}}
\end{align*}
satisfy $\displaystyle \lim_{N\rightarrow\infty}A^N_t\!=\!A_t$ a.s.\
and $\displaystyle \lim_{N\rightarrow\infty}b^N_t\!=\!b_t$ a.s.\footnote{
	The algorithms were designed so that these weakly-dependent SLLN's hold.
	However, there is future mathematical work
	to be done along the lines of \cite{BranchRates} to obtain theoretical convergence results.}, where
\begin{equation*}
A_t=\frac{E[ L_te^J(S_t,V_t)e^J(S_t,V_t)'1_{\{Z_t>0\}}]} {P(Z_t>0)}=\!\frac{\widetilde E[e^J(S_t,V_t)e^J(S_t,V_t)'1_{\{Z_t>0\}}]}{P(Z_t>0)}
\end{equation*}
and
\begin{equation*}
b_t=\frac{E[ L_tZ_{\tau^{J}_{t+1}}e^J(S_t,V_t)1_{\{Z_t>0\}}]}{P(Z_t>0)}=\!\frac{\widetilde E[Z_{\tau^{J}_{t+1}}e^J(S_t,V_t)1_{\{Z_t>0\}}]}{P(Z_t>0)},
\end{equation*}
with $\frac{d\widetilde P}{dP}\bigg|_{\mathcal F_t}=L_t$.
Now, suppose $\alpha^{J,j}_t$ is defined by
$\alpha^{J,0}_t=0$ and $k=1$, and by
\begin{align*}
(\alpha^{J,j}_t,k)&=
\begin{cases}
(\alpha^{J,j-1}_t,k),&Z_t^j=0\\
\alpha^{J,j-1}_t+\frac{\gamma L_t^j}{k^\chi}( Z^j_{\tau^{J,j}_{t+1}} -e^J(S^j_t,V^j_t)' \alpha^{J,j-1}_t)e^J(S^j_t,V^j_t),k+1),&Z_t^j>0
\end{cases}\\
&=
\begin{cases}
(\alpha^{J,j-1}_t,k),&Z_t^j=0\\
(\alpha^{J,j-1}_t+\frac{\gamma}{k^\chi}\frac{\sum_{i=1}^N1_{Z_t^j>0}}N( b_j -A_j \alpha^{J,j-1}_t),k+1),&Z_t^j>0,
\end{cases}
\end{align*}
for $j=1,2,...,N$.
Then, as explained in \cite{Kouritzin16}
$\displaystyle \lim_{N\rightarrow\infty}\alpha^{J,N}_t=\alpha^J_t=A_t^{-1}b_t$ 
a.s.\ for any $\gamma>0$ and 
\begin{equation*}
\lim_{N\rightarrow\infty}\frac1N \sum_{j=1}^N Z_{\tau^{J,j}_0}^j=\widetilde E[Z_{\tau^J_0}],	
\end{equation*}
which with help of \cite{ClPr} yields
\begin{equation*}
\lim_{J\rightarrow\infty}\lim_{N\rightarrow\infty}\frac1N \sum_{j=1}^N Z_{\tau^{J,j}_0}^j=\widetilde E[Z_{\tau_0}]	
\end{equation*}
and gives a means to estimate the option value $\widetilde E[Z_{\tau_0}]$ as well as optimal execution $\tau_0$. This procedure is described in Algorithm \ref{algo:SA}.

Algorithm \ref{algo:SA} can be improved by using the average of the coefficients ${\bar\alpha^{J,N}_t = \frac 1N \sum_{j=1}^N \alpha^{J,j}_t}$ to approximate $\widetilde E[Z_{\tau_t^J}]$. This modification is done by initializing $\bar\alpha_t^J=0$ for all $t \in \{0,\ldots,T-1\}$, by adding the following line after Line \ref{line:alphat}:
\begin{equation*}\label{eq:ave_alpha}
\bar{\alpha}_t^{J} = ((k-1)\bar{\alpha}_t^{J}  + \alpha_t^{J})/k,
\end{equation*}
and by calculating the projection in line \ref{line:projection} using $e^J(S^j_t,V^j_t)^\prime \bar\alpha^{J}_t$. 
Generally, the exponent $\chi$ in the gain step $\frac{\gamma}{k^\chi}$ is equal to 1 when there is no averaging. 
However, the best value for $\chi$ is generally below $1$ with averaging. 
With the right parameters, averaging the coefficients speeds up convergence of the algorithm.
Another important advantage of this modification is that it significantly increases the robustness of the algorithm with respect to the parameters $\gamma$ and $\chi$; the new algorithm is thus easier to use in practice because convergence is achieved for wider intervals of parameters.

There is still is a potential problem since the projection coefficients
depend upon the stopping times and the stopping times depend upon
the projection coefficients.
Fortunately, the dependence can be decoupled: $\alpha^{J,j}_t$ depends
upon $\tau_{t+1}^{J,j}$ and $\tau_{t}^{J,j}$ depends upon $\alpha_t^{J,N}$,
which means that we have to work backwards, use the fact $\tau_T^{J,j}=T$
and compute $\alpha_t^{J,N}$ prior to $\tau_{t}^{J,j}$.
This is reflected in the following algorithm, in which $\{e_k\}_{k=1}^J$ are the chosen basis functions and $\gamma$ is a positive constant.
\begin{algorithm}[h!]
	\caption{Stochastic Approximation}\label{algo:SA}
	\begin{algorithmic}[1]
		\Procedure{AmericanOptionPricing}{$e_k$, $\gamma$, $\chi$, $N$, $T$}
		\State $\zeta = 0$, $\lambda = 0$
		\State $\{\alpha^J_t = 0\}_{t=0}^{T-1}$, $\{\tau^{J,j} = T\}_{j=1}^N$
		\State Simulate $\{L^j,S^j,V^j,R^j\}_{j=1}^N$, copies of $(\hat{L},\hat{S},\hat{V},R)$\footnotemark
		\State $Z^j_t=p(t,S^j_t)$ or $p(t,R^j_t)$ for all $t \in \{0,\ldots,T\}$, $j \in \{1,\ldots,N\}$. 
		\For{$t$}{$T-1$}{$0$}
		\State $k=0$
		\For{j}{1}{N}
		\If{$Z^j_t > 0$}
		\State $k = k+1$
		\State $\alpha^{J}_t=\alpha^{J}_t+\frac{\gamma L_t^j}{k^\chi}( Z^j_{\tau^{J,j}} -e^J(S^j_t,V^j_t)' \alpha^{J}_t)e^J(S^j_t,V^j_t)$
		\label{line:alphat}
		\EndIf
		\EndFor
		\For{$j$}{1}{$N$}
		\If{$Z^j_t > 0$ and $Z_t^j\ge e^J(S^j_t,V^j_t)^\prime\alpha^{J}_t$}
		\label{line:projection}
		\State $\tau^{J,j}=t$
		\EndIf
		\EndFor		
		\EndFor
		\State Option value = $\frac{\sum_{j=1}^N L^j_{\tau^{J,j}} Z^j_{\tau^{J,j}}}{\sum_{j=1}^N Z^j_{\tau^{J,j}}}$
		\EndProcedure
	\end{algorithmic}
\end{algorithm}
\footnotetext{$R^j$ only need to be simulated when pricing Asian options with early exercise.}

\section{Numerical Results} \label{EmpResults}

In this section, we illustrate how branching can be used to increase the accuracy and efficiency of Algorithm \ref{algo:Weighted}. 
This algorithm was shown in \cite{Kouritzin16} to be an interesting tool to price path-dependent options in the Heston model. 
However, a disadvantage of the simulation method is that it relies on likelihood, or weights, whose variance increase in time, possibly reducing the accuracy of Monte Carlo price estimators. 

Here we consider three different parameter sets to identify cases where branching clearly improves the performance of the weighted simulation algorithm.
The parameters considered are given in Table \ref{Tab:Market parameters}.\footnote{The parameters were chosen to illustrate the efficiency of the branching algorithms rather than for their fit to a specific set of financial data.}
We show that, for a given number of simulations, branching is often very effective in increasing the precision of price estimates and in decreasing their variance.

To perform the simulations, we discretize the time horizon using $m$ time steps per year. 
We first consider both Asian and European straddle options with maturity $T$, thus using $Z_T = p(T,S_T) = |S_T-K|$ for European options and $Z_T = p(T,R_T) = |R_T-K|$ for Asian options.

\begin{table}[h!]
	\centering
	\small
	\caption{Market parameters used for numerical examples.}
	\label{Tab:Market parameters}
	\begin{tabularx}{\textwidth}{*{4}{Y}}
		\toprule
		~ & PS1 & PS2 & PS3 \\
		\cmidrule{2-4}
		$S_0$ & 100 & 100 & 100 \\
		$\mu$ & 0.02 & 0.02 & 0.02 \\
		$\nu$ & 0.085 & 0.424 & 0.225 \\
		$\varrho$ & 6.21 & 6.00 & 2.86 \\
		$\kappa$ & 0.2 & 0.8 & 0.6 \\
		$\rho$ & -0.7 & -0.75 & -0.96 \\
		$V_0$ & 0.501 & 0.11 & 0.07 \\
		\cmidrule{2-4}
		$n$ & 9 & 3 & 3 \\
		${4\nu}/{\kappa^2}$ & 8.50 & 2.65 & 2.50\\
		\bottomrule
	\end{tabularx}
\end{table}

In all cases, the use of the explicit weak solution of Theorem \ref{Theorem2} necessitates the simulation of weights $L^j$ since $\frac{4\nu}{\kappa^2}$ is not an integer. 

We first price options without early exercise to highlight the effect of branching on the efficiency of the simulation algorithm. 
We then price American-type options using the algorithm of Section \ref{Pricing}  and show that when re-sampling is needed, branching algorithms should be used instead of the bootstrap algorithm.

\subsection{Efficiency of the Simulation Algorithms}\label{ssec:european_numerical_results}

In this section, we price financial options using the branching algorithms introduced in Section \ref{BranAlg} in combination with the weighted simulation method (Algorithm \ref{algo:Weighted}). The Monte Carlo price estimates are calculated using the weighted simulated price paths and we assume that no early exercise is allowed. In other words, the price estimate of an option with payoff $Z_T$ is given by 

\begin{equation}
\widehat{C} = \frac{\sum_{j=1}^N L^{j}_T Z^{j}_T}{\sum_{j=1}^N L^{j}_T}.\label{eq:weightedMCprice}
\end{equation}

Simulations are obtained using Algorithm \ref{algo:Weighted} with Simpson's $\frac13$ rule with $50$ time steps per year and $M$ sub-intervals (see Table \ref{Tab:Simulation parameters} for specific values) to compute the deterministic integrals. The branching parameters $r_t$ (for combined branching) and $c^{eff}$, $c^{neff}$ (for effective branching), and the stopping parameter $\varepsilon$ were chosen in the following way:
\begin{enumerate}
	\item 
	Find $\varepsilon^*$ minimizing $\left|\frac1N\sum_{j=1}^N(L^j_T-1)\right|$.
	\item 
	Using $\varepsilon^*$, find  $r_t$ (for combined branching) or $c^{eff}$, $c^{neff}$ (for effective branching) minimizing $\left|\frac{\sum_{j=1}^N(L^j_TS^j_T-e^{\mu T}S_0)}{\sum_{j=1}^N L^j_T}\right|$.
\end{enumerate}
Using the known quantities $E[L_T]=1$ and $E[S_T]=e^{\mu T}S_0$ allows for efficient calibration of the branching parameters. We note however that, depending on the problem, the objective function $\left|\frac{\sum_{j=1}^N(L^j_TS^j_T-e^{\mu T}S_0)}{\sum_{j=1}^N L^j_T}\right|$ is not smooth, since it is based on simulations, and often almost flat around its minimum, which makes it more difficult to identify optimal parameters. However, it is possible to identify intervals of good parameters that can be used in the algorithms.

We note that in the cases of PS2 and PS3, $\frac{4\nu}{\kappa^2}$ is close to 2, which is the threshold for the Feller condition. Thus, while the Feller condition ensuring that $V_t$ does not hit 0 is satisfied in all cases, we expect $V_t$ to drop much closer to 0 when using PS2 and PS3, compared to PS1. Since the explicit solution underlying the weighted simulation algorithm only holds until $V_t$ gets arbitrarily close to 0, parameter sets that keep $V_t$ away from 0 should be ``easier'' to simulate from. This explains why for the second and the third parameter set, we need smaller time sub-intervals to simulate $V_t$ (that is, $M=6$) and a higher $\varepsilon$. Indeed, setting $\varepsilon = 10^{-5}$ helps catch the few paths where the volatility drops too much, causing the weights to explode. When this happens, the weight is set to 0; the particle is effectively dropped from the Monte Carlo estimate, inducing a bias, which we show to be small.

\begin{table}
	\centering
\small
\caption{Simulation parameters used for numerical examples.}
\label{Tab:Simulation parameters}
\begin{tabularx}{\textwidth}{*{4}{Y}}
	\toprule
	~ & PS1 & PS2 & PS3 \\
	\cmidrule{2-4}
	$M$ & 2 & 6 & 6 \\
	$m$ & 50 & 50 & 50 \\
	$r_t$ & 1.05 & 1.05 & 1.05\\
	$c^{eff}$ & 1.055 & 1.05 & 1.045\\
	$c^{neff}$ & 0.2 & 2 & 1.5 \\
	$\varepsilon$ & $10^{-10}$ & $10^{-5}$ & $10^{-5}$\\
	\bottomrule
\end{tabularx}
\end{table}

\subsubsection{Accuracy of the Price Estimate}

There exists situations where branching is crucial to obtaining a fast and accurate pricing algorithm. We illustrate this situation by pricing a European straddle option (for which a closed form is available) using the market parameters of Table \ref{Tab:Market parameters} and the simulation parameters of Table \ref{Tab:Simulation parameters}. That is, we calculate the Monte Carlo estimator of the option price $\widehat{C}$ using \eqref{eq:weightedMCprice}.

To test the accuracy of the various pricing algorithms, we use the relative RMSE
\begin{equation*}
\frac{\sqrt{E[(\widehat{C}-C)^2]}}{C},
\end{equation*}
where $C$ is the real option price calculated using the integral form available for European option prices in the Heston model (see \cite{He} and \cite{albrecher2007little}). 
We compute the relative RMSE for $N \in \{10^4, 5 \times 10^4, 10^5, 5 \times 10^5\}$ simulations.

\begin{figure}
	\centering
	\begin{subfigure}[b]{0.45\textwidth}
		\includegraphics[width=\textwidth]{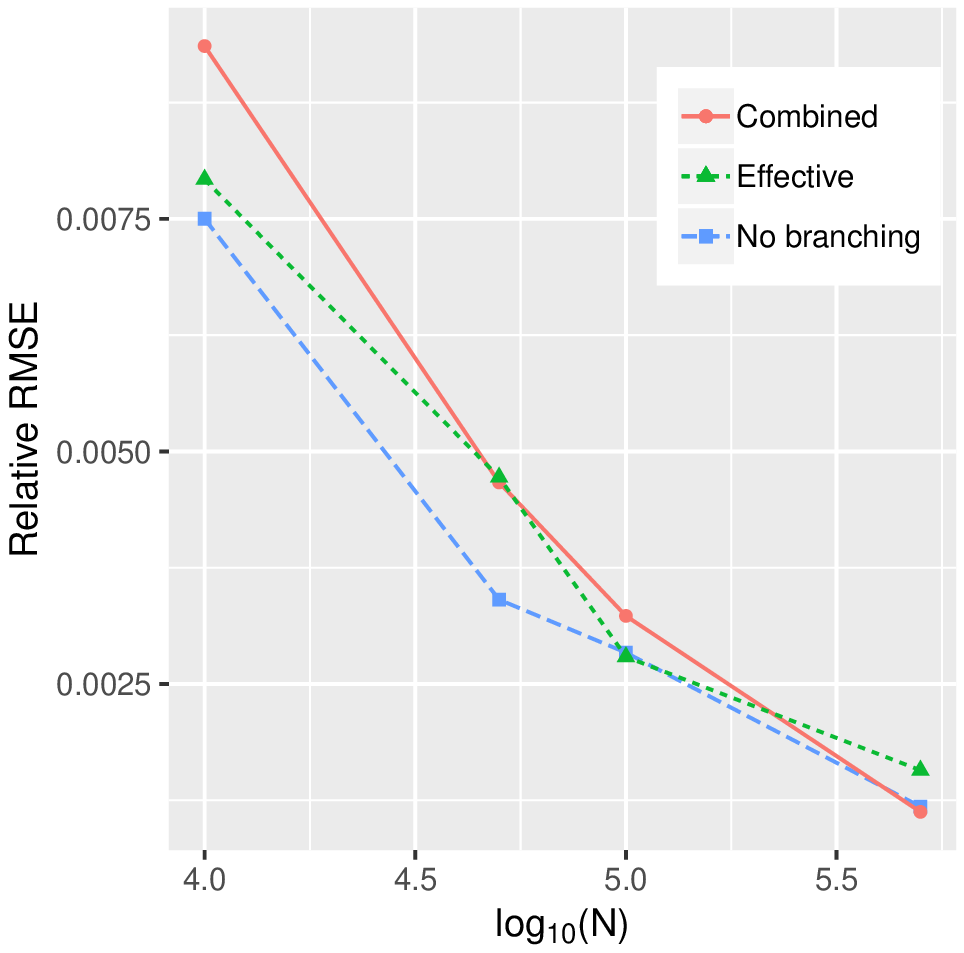}
		\caption{PS1}
		\label{fig:PS1_RMSE}
	\end{subfigure}
	~ 
	\begin{subfigure}[b]{0.45\textwidth}
		\includegraphics[width=\textwidth]{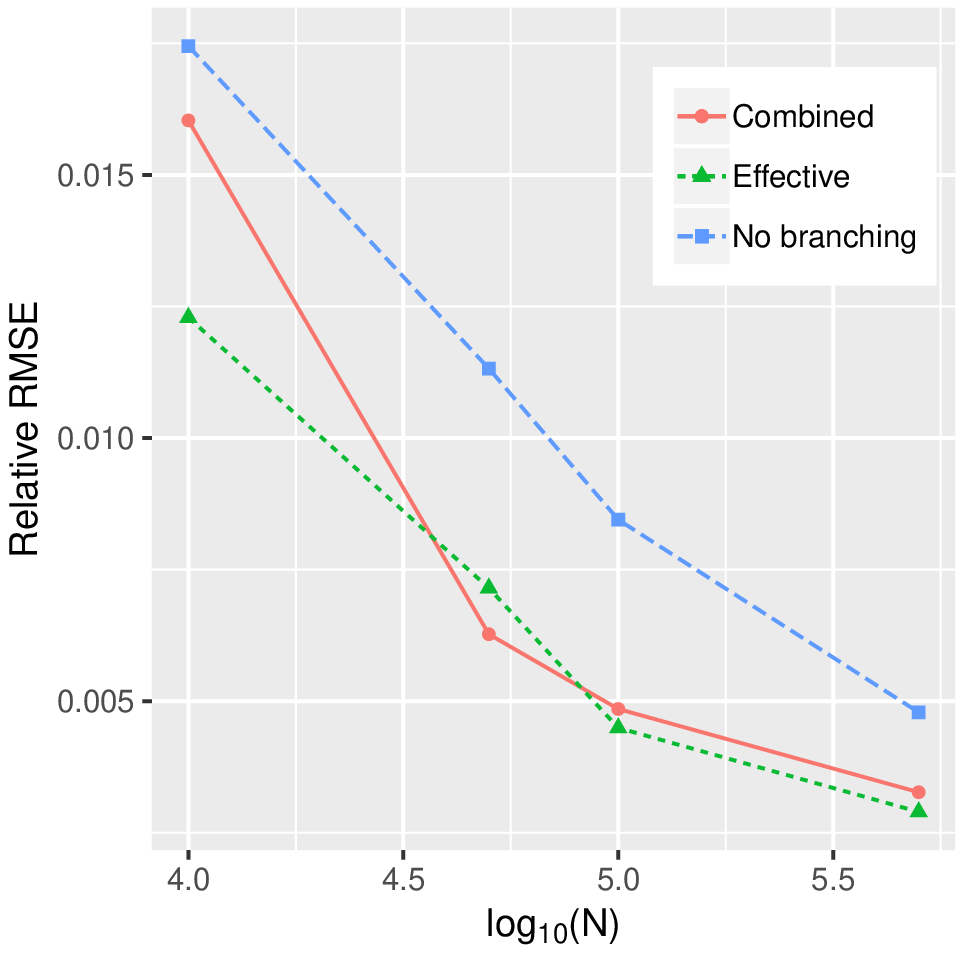}
		\caption{PS2}
		\label{fig:PS2_RMSE}
	\end{subfigure}\\
	~ 
	\begin{subfigure}[b]{0.45\textwidth}
		\includegraphics[width=\textwidth]{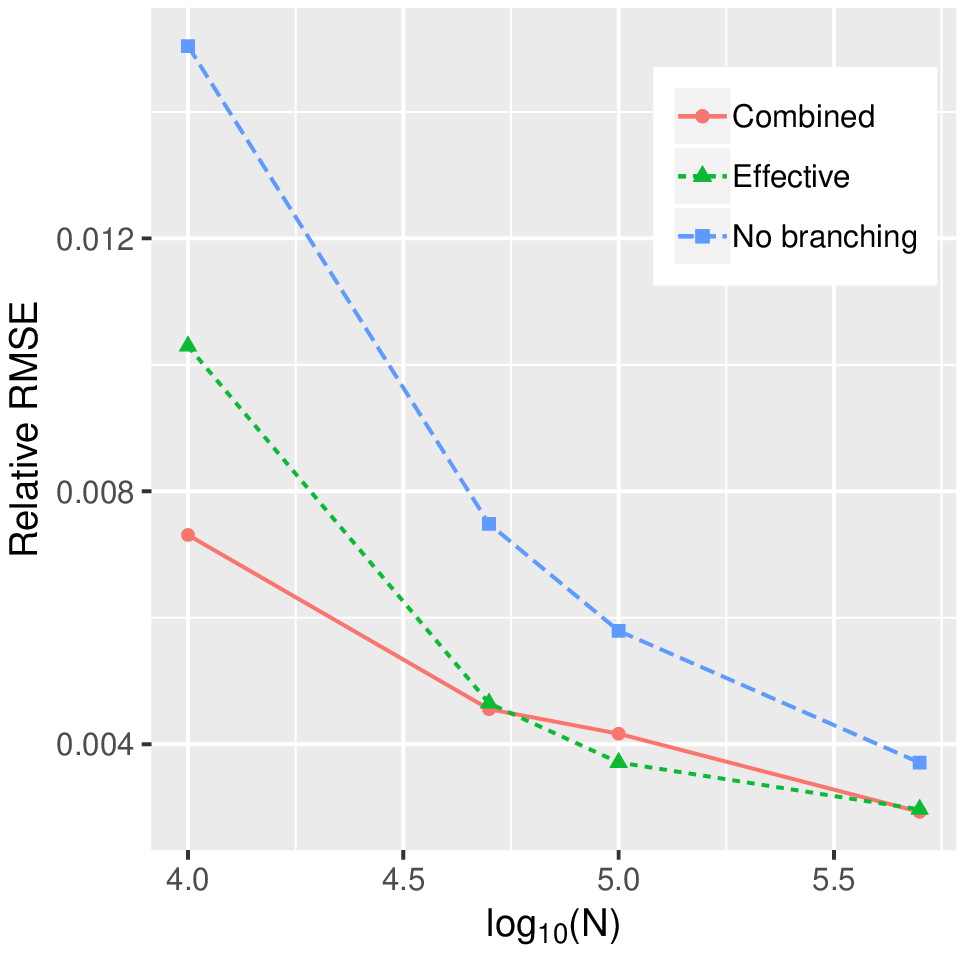}
		\caption{PS3}
		\label{fig:PS3_RMSE}
	\end{subfigure}
	\caption{Relative RMSE of the price estimator $\hat{C}$ in terms of number of simulations, European straddle options.}\label{fig:RMSE}
\end{figure}

Figure \ref{fig:RMSE} shows that, for the second and third parameter sets (PS2 and PS3), the relative RMSE of the price estimate is significantly reduced by branching. Branching is however less necessary for the first parameter set (PS1), for which the weighted algorithm \ref{algo:Weighted} alone results in very precise estimates.

As mentioned above, the branching parameters $r_t$, $c^{eff}$ and $c^{neff}$ were chosen to optimize the performance of the pricing algorithm. Depending on the parameters used, the resulting optimal proportion of branched particles vary. This proportion is much lower for PS1 (under 20\%) than for PS2 and PS3 (between 30\% and 35\%).

The choice of $M$, the number of intervals used to approximate the numerical integrals, depends on the parameters. Table \ref{Tab:M_sens} shows the sensitivity of the RMSE of the price estimate to changes in $M$. To isolate the impact of $M$ on the RMSE, we perform all calculations without resampling the particles. The results obtained with PS1 are already very precise with $M=2$; increasing it does not reduce the RMSE, but slows down the algorithm. This is not true for PS2 and PS3; increasing $M$ has a significant impact of the RMSE. We note that in the case of PS2, using $M=8$ could lead to more precise results, while also further increasing the execution time. We also observe that for a fixed $M$, the algorithm is much slower for PS1, since it requires the simulation of $n=9$ Gaussian processes (compared to $n=3$ for PS2 and PS3). However, using $M=2$ allows the run time to remain low.

\begin{table}[h!]
	\centering
	\caption{Run times (in seconds) and relative RMSE of the price estimator $\hat{C}$ as a function of $M$, European straddle options, $N=10^5$ simulations, no branching.}
	\label{Tab:M_sens}
	\begin{tabularx}{\textwidth}{*{5}{Y}}
		\toprule
		$M$ & 2 & 4 & 6 & 8 \\
		\cmidrule{1-5}
		& \multicolumn{4}{c}{PS1}\\
		\cmidrule{1-5}
		RMSE & 0.0017 & 0.0022 & 0.0026 & 0.0022\\
		Run time & 3.98 & 7.14 & 10.32 & 13.50\\
		\cmidrule{1-5}
		& \multicolumn{4}{c}{PS2}\\
		\cmidrule{1-5}
		RMSE & 0.4223 & 0.0315 & 0.0177 & 0.0044\\
		Run time & 1.90 & 2.99 & 4.09 & 5.21\\
		\cmidrule{1-5}
		& \multicolumn{4}{c}{PS3}\\
		\cmidrule{1-5}
		RMSE & 0.2054 & 0.0105 & 0.0047 & 0.0050\\
		Run time & 1.90 & 2.98 & 4.09 & 5.21\\
		\bottomrule
	\end{tabularx}
\end{table}

\subsubsection{Performance of the Branching Algorithms}

We have shown that for a given number of initial particles, branching can increase the accuracy of the price estimate. This additional step can however slow down the pricing algorithm. In this section, we consider the standard deviation of the price estimate as a function of time. In addition to European options, we also price Asian options with fixed strike $K=100$, for which there does not exist a closed-form expression. Because of its speed, the weighted simulation algorithm is particularly well-suited for path-dependent options (see also \cite{Kouritzin16}).

To compare the performance of the algorithms, we consider the standard deviation of the estimate relative to the true option price, which we approximate by computing the price estimate 50 times. The true option price is computed using the semi-analytical formula in the case of the European option. The mean of 50 price estimates obtained using $N = 5 \times 10^5$ and the weighted algorithm with effective particle branching is considered to be the ``true'' price of the Asian option. 

\begin{figure}
	\centering
	\begin{subfigure}[b]{0.45\textwidth}
		\includegraphics[width=\textwidth]{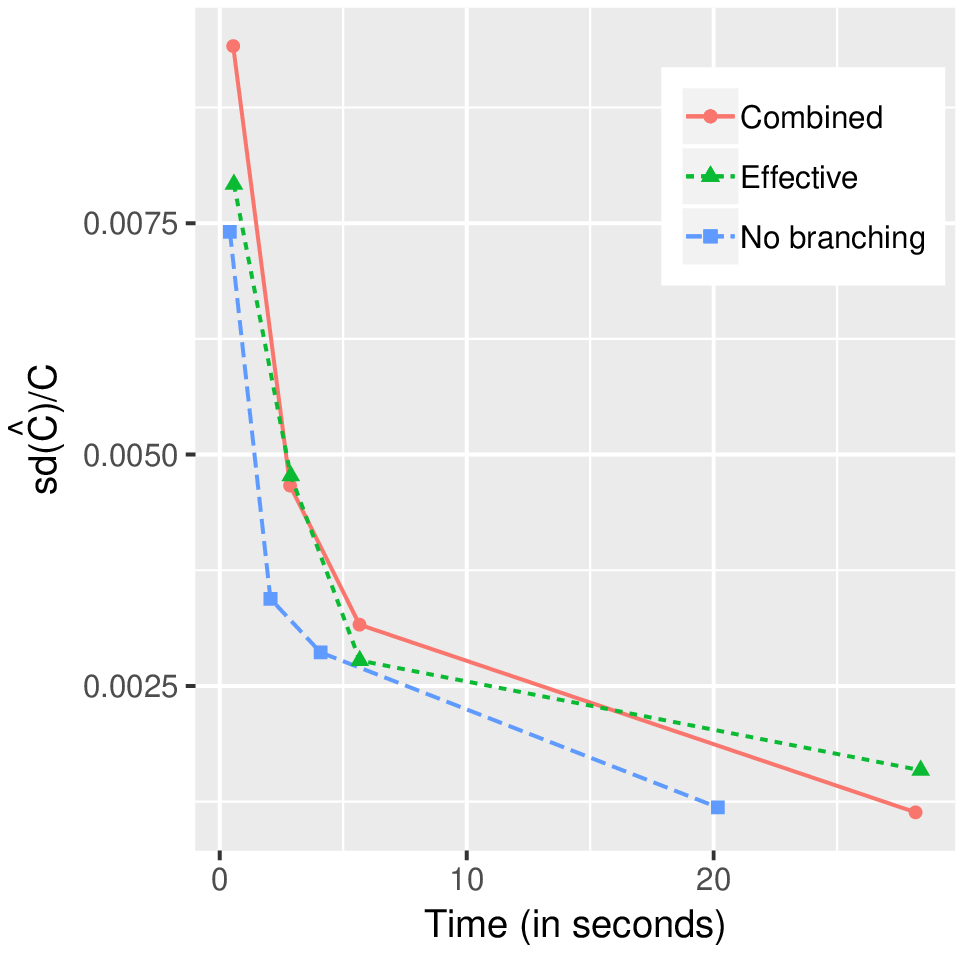}
		\caption{PS1}
		\label{fig:PS1_sdEuro}
	\end{subfigure}
	~ 
	\begin{subfigure}[b]{0.45\textwidth}
		\includegraphics[width=\textwidth]{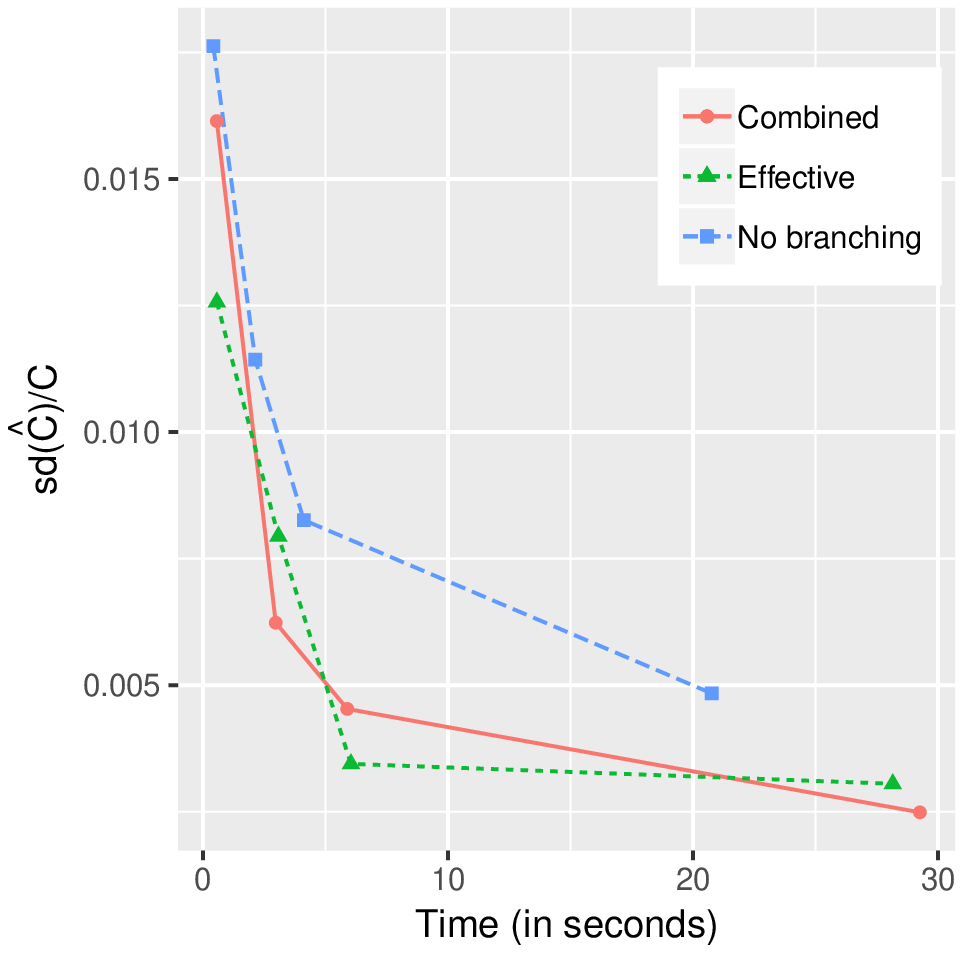}
		\caption{PS2}
		\label{fig:PS2_sdEuro}
	\end{subfigure}\\
	~ 
	\begin{subfigure}[b]{0.45\textwidth}
		\includegraphics[width=\textwidth]{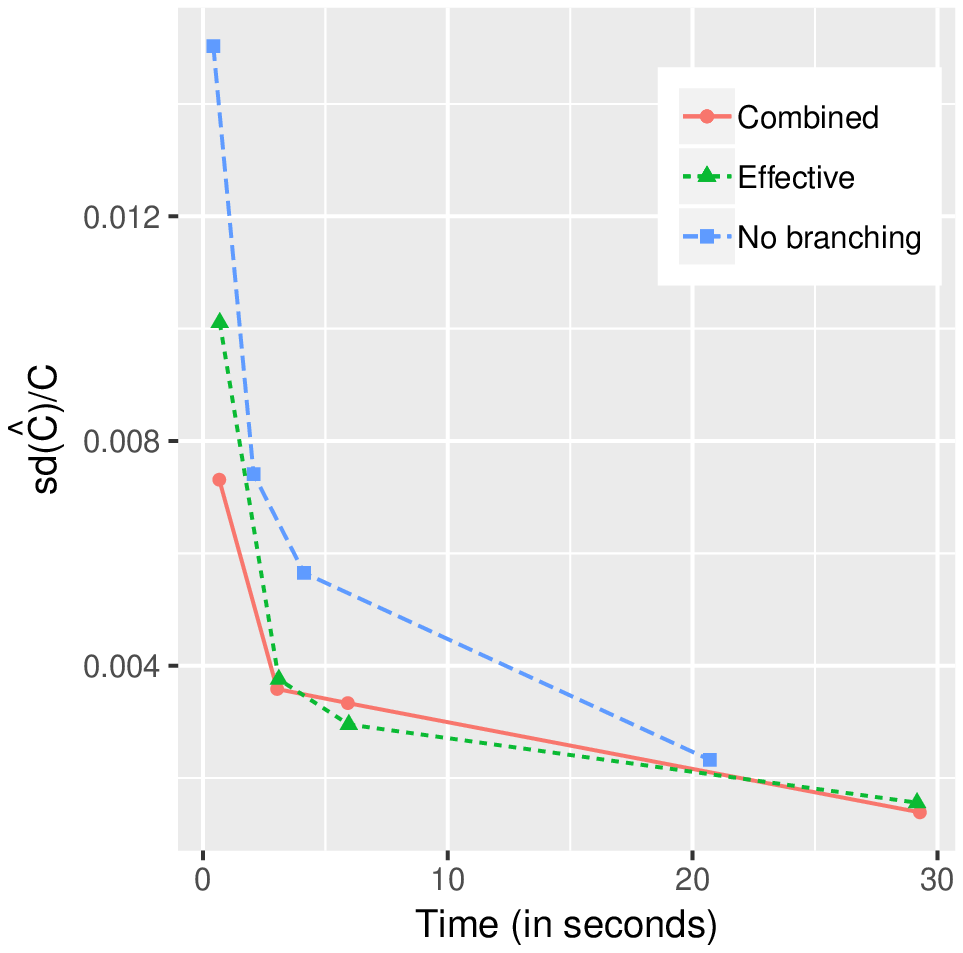}
		\caption{PS3}
		\label{fig:PS3_sdEuro}
	\end{subfigure}
	\caption{Relative standard deviation of the price estimator $\widehat{C}$ in terms of number of simulations, European straddle options.}\label{fig:sdEuro}
\end{figure}

\begin{figure}
	\centering
	\begin{subfigure}[b]{0.45\textwidth}
		\includegraphics[width=\textwidth]{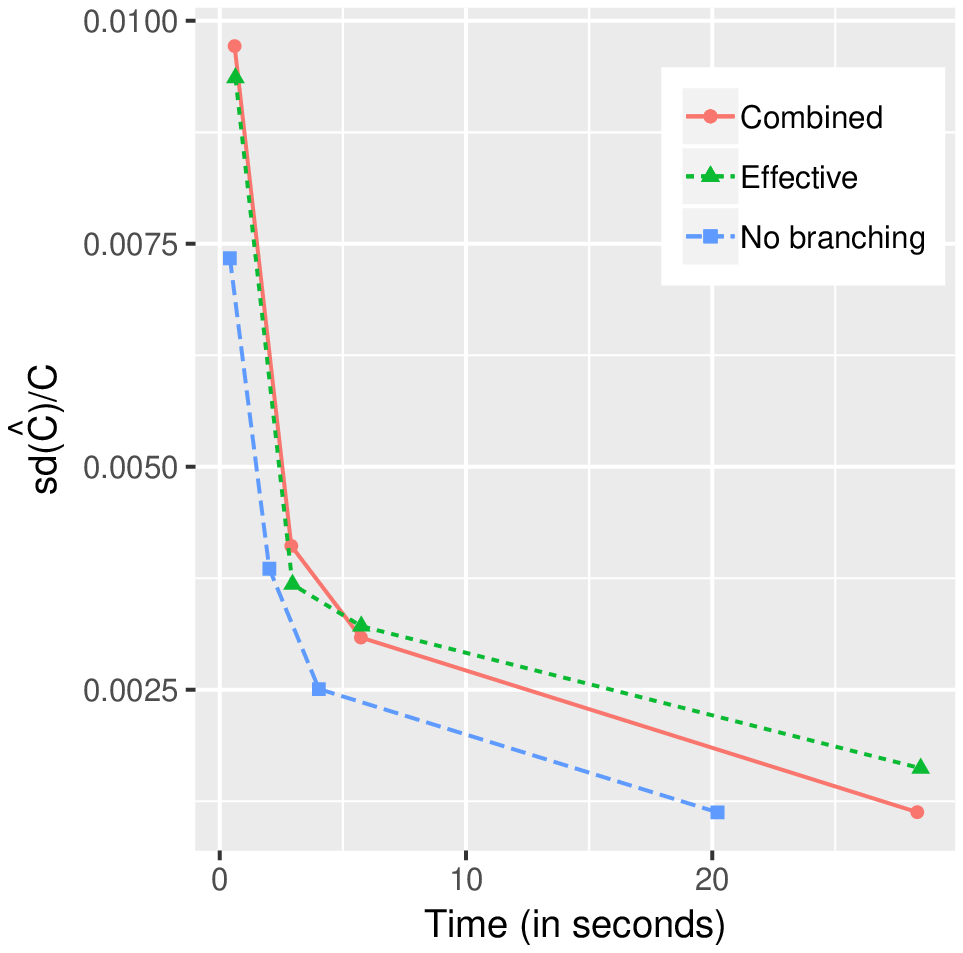}
		\caption{PS1}
		\label{fig:PS1_sdAsian}
	\end{subfigure}
	~ 
	\begin{subfigure}[b]{0.45\textwidth}
		\includegraphics[width=\textwidth]{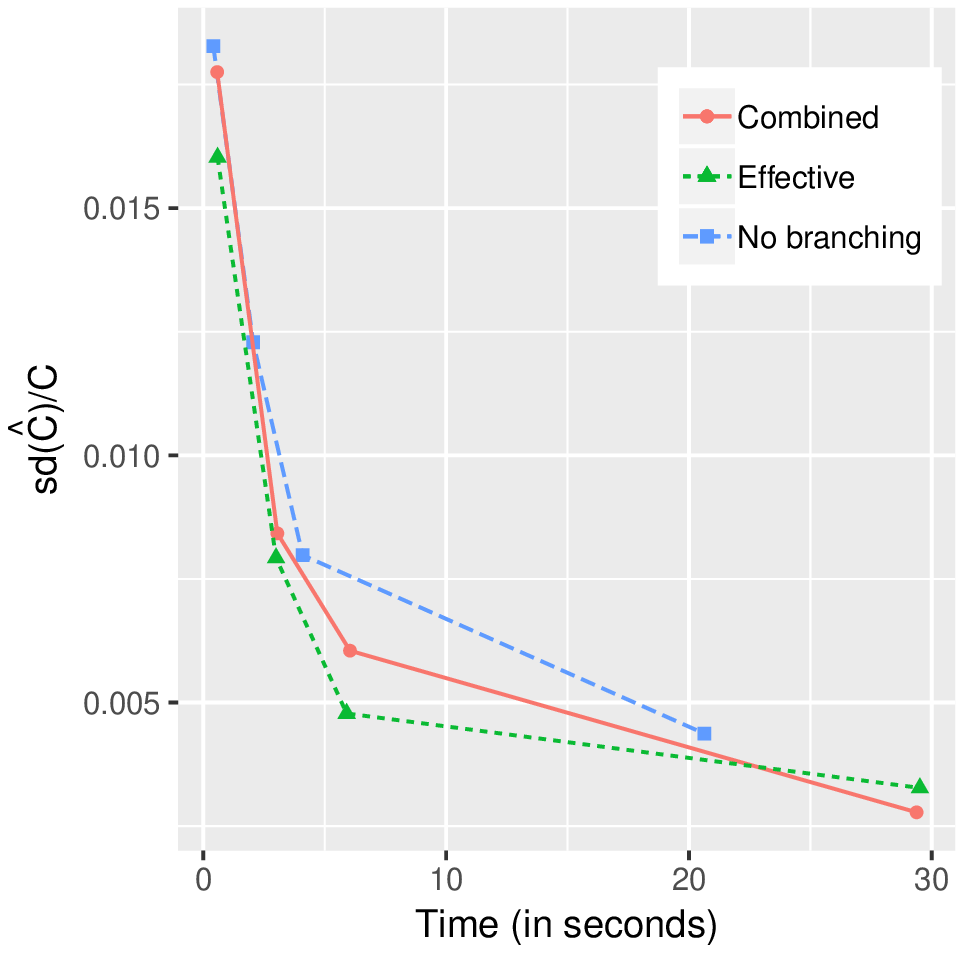}
		\caption{PS2}
		\label{fig:PS2_sdAsian}
	\end{subfigure}\\
	~ 
	\begin{subfigure}[b]{0.45\textwidth}
		\includegraphics[width=\textwidth]{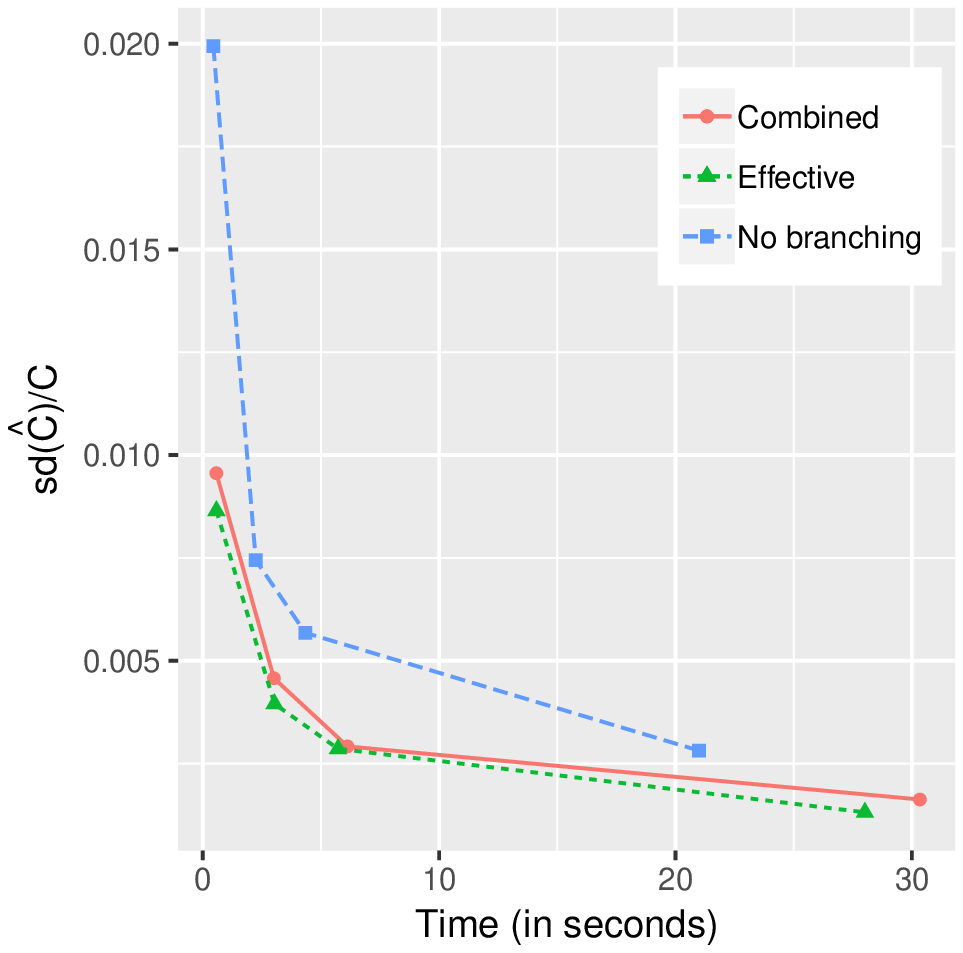}
		\caption{PS3}
		\label{fig:PS3_sdAsian}
	\end{subfigure}
	\caption{Relative standard deviation of the price estimator $\widehat{C}$ in terms of number of simulations, Asian straddle options.}\label{fig:sdAsian}
\end{figure}

There are cases in which branching clearly increases the performance of the pricing algorithm. This is true for PS2 and PS3; it can be seen in Figures \ref{fig:sdEuro} and \ref{fig:sdAsian} that the standard deviation of the price estimator can be reduced much faster when branching is used. That is, although branching does slow down the pricing procedure, the resulting improvement is such that fewer initial particles are needed for the standard deviation of the price estimate to reach a level similar to the one reached without branching for a larger number of initial particles. For example, in the case of PS2 (see Figure \ref{fig:PS2_sdEuro}), the European price estimate using either branching algorithms and $N=10^5$ has a smaller standard deviation than the non-branched estimate obtained with five times more simulated paths. In this case, pricing with branching is more than 3 times faster.  Significant improvements due to branching are also observed for the third parameter set and when pricing Asian options (see Figure \ref{fig:sdAsian}).

In the case of PS1, the algorithm is already very precise without branching; additional resampling does not seem to be necessary. However, it does not significantly increase the RMSE or the standard deviation of the European option price estimate. Branching does however increase the standard deviation of the Asian option price estimate. This is most likely due to the path-dependent nature of the Asian option payoff; branching can lead to path degeneracy (the surviving particles having few distinct ancestors), which impacts the variance of the price estimate. Although path degeneracy does not appear be a significant problem in our numerical examples in general, the very high performance of the weighted simulation algorithm (without resampling) for PS1 leads to the variance of the price estimate increasing when resampling is added. 

The need for resampling is linked to the size of the ratio $\frac{4\nu}{\kappa^2}$. Indeed, this ratio is much higher for PS1 ($\frac{4\nu}{\kappa^2}=8.50$); we could say that the Feller condition is ``completely satisfied'' and $V_t$ stays well away from 0. Further numerical tests also show that the variance of the likelihood $L_T$ is much lower than in the case of the other two parameter sets. PS2 and PS3 have much smaller ratios $\frac{4\nu}{\kappa^2}$; they are equal to 2.65 and 2.50, respectively. For such parameters, $V_t$ has a higher probability of dropping closer to 0, which increases the variance of the likelihoods and creates the need for more branching.

Our numerical results show that in general, when Algorithm \ref{algo:Weighted} is used, a branching algorithm should be implemented alongside. Effective branching, of which combined branching is a special case, should be prioritized. When necessary, depending on the market parameters considered, branching can be turned off by setting $c^{eff}$ and $c^{neff}$ high enough such that no particle is branched. Improvements due to branching are most significant when the variance of the weights $L$ is high; for example, when $\frac{4\nu}{\kappa^2}$ is close to 2 (as with PS2 and PS3) or when the option to price has a longer maturity, since the variance of the particle system tends to increase with time.

\begin{remark}
	In our setting, the performance of the combined branching algorithm is similar to the one of the effective particle branching one. 
	This is likely due to the fact that, in the cases we study, the number of effective particles remains relatively constant as $t$ increases. 
	Thus, using both algorithms results in resampling a similar proportion of the particles at each time step, whether $r_t$ is kept constant (combined branching) or not (effective particle branching). 
	In situations where the distribution of the weights can vary widely from one time step to the other (in particle filtering, for example), the increased flexibility provided by the effective particle branching algorithm can result in a significant performance improvement (see \cite{BranchDraft}).
	Since it is similar to the combined branching algorithm in terms of complexity of implementation and computation ressources, we still recommend the use of the effective particle branching algorithm, even in cases where its advantage is less obvious.
\end{remark}

\subsection{Comparison with Bootstrap Algorithm}\label{ssec:asian_numerical_results}

In this section, we highlight an important advantage of the branching algorithms of Section \ref{BranAlg} compared to the bootstrap algorithm of Section \ref{sec:bootstrap}: the preservation of the process distribution. 
To do so, we price Asian  options with early exercise, whose price strongly depends on the distribution of the historical paths.
We focus on the second and the third sets of parameters (PS2 and PS3) since they are the ones for which the performance of the pricing algorithm is the most improved by resampling.

To price the options, we use the SA/DP algorithm exposed in Section \ref{Pricing} using the first $\sqrt[3]{J}$ weighted Laguerre polynomials as basis functions for each of volatility, price and average price.
We also use the averaged coefficients $\bar{\alpha}_t^{J,N}$ in order to increase the robustness of the algorithm to the choice of parameters $\gamma$ and $\chi$.
In our setting, compared to the well-known least-squares Monte Carlo algorithm, SA/DP allows us to use a large number of basis functions without sacrificing numerical stability and at a lower computing cost than the least-square Monte Carlo algorithm. This leads to more precise price estimates (see \cite{Kouritzin16}).

As explained in Section \ref{sec:simulation}, the amount of noise introduced by bootstrap resampling can affect the historical law of the particle system, which is problematic when pricing options with early exercise via recursive conditional expectations determined from cross-sectional data, as in Algorithm \ref{algo:SA}.
We price at-the-money Asian call options with fixed strike and early exercise (that is, we consider Bermudan-style options that can be exercised at every discrete time step). 
To do so, we use Algorithm \ref{algo:SA} with $Z^j_t = \max(R^j_t-K,0)$ and $K=S_0$.

\begin{figure}
	\centering
	\begin{subfigure}[b]{0.45\textwidth}
		\includegraphics[width=\textwidth]{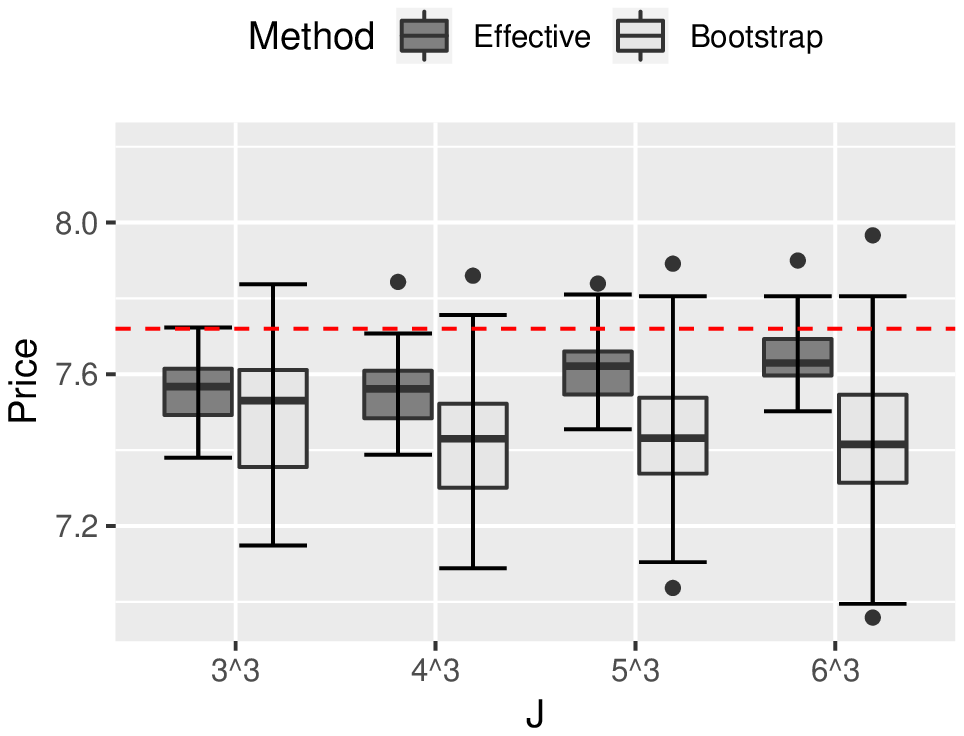}
		\caption{$N=10^5$}
		\label{fig:PS2_ame_100}
	\end{subfigure}
	~ 
	\begin{subfigure}[b]{0.45\textwidth}
		\includegraphics[width=\textwidth]{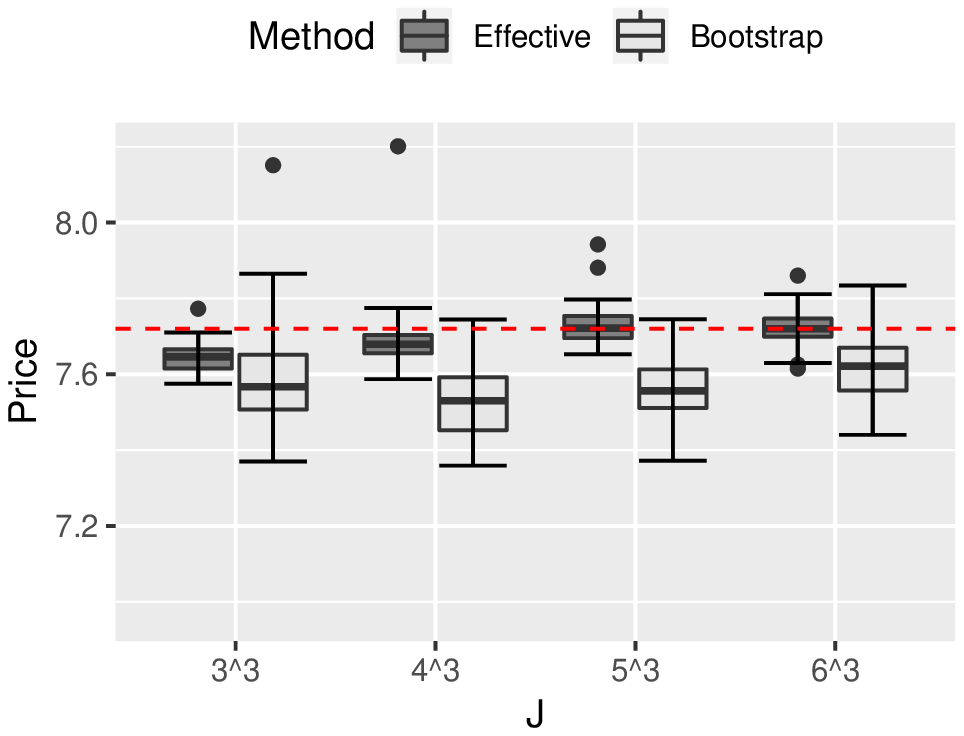}
		\caption{$N=5 \times 10^5$}
		\label{fig:PS2_ame_500}
	\end{subfigure}
	\caption{Price estimates of Asian options with early exercise (PS2), true price (7.67) indicated by the dashed line.}
	\label{fig:PS2_ame}
\end{figure}

\begin{figure}
	\centering
	\begin{subfigure}[b]{0.45\textwidth}
		\includegraphics[width=\textwidth]{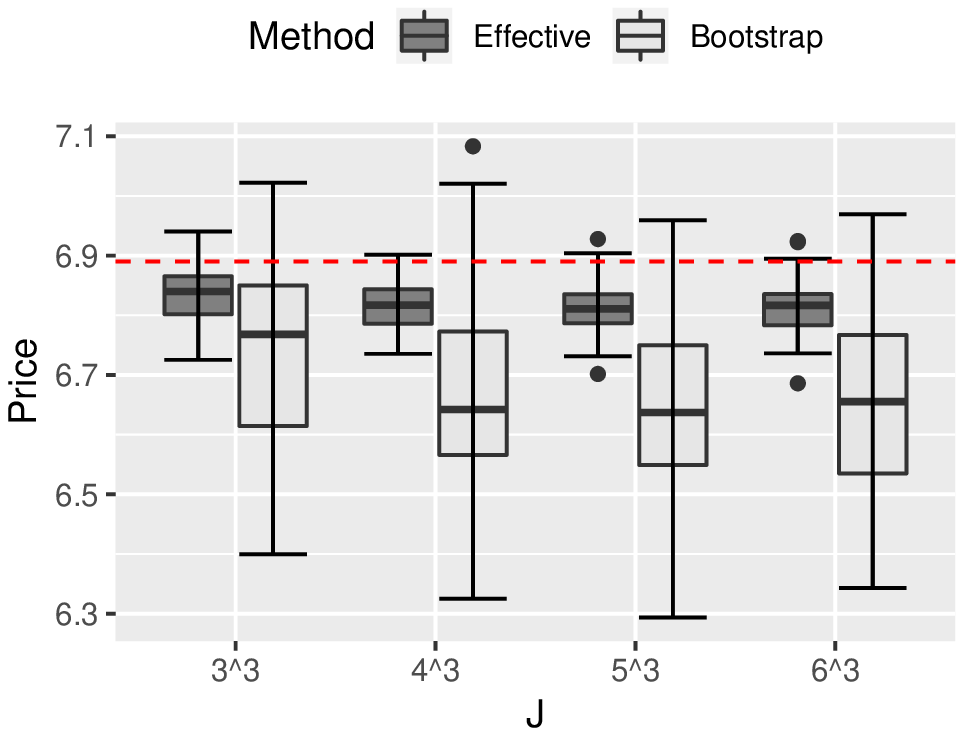}
		\caption{$N=10^5$}
		\label{fig:PS3_ame_100}
	\end{subfigure}
	~
	\begin{subfigure}[b]{0.45\textwidth}
		\includegraphics[width=\textwidth]{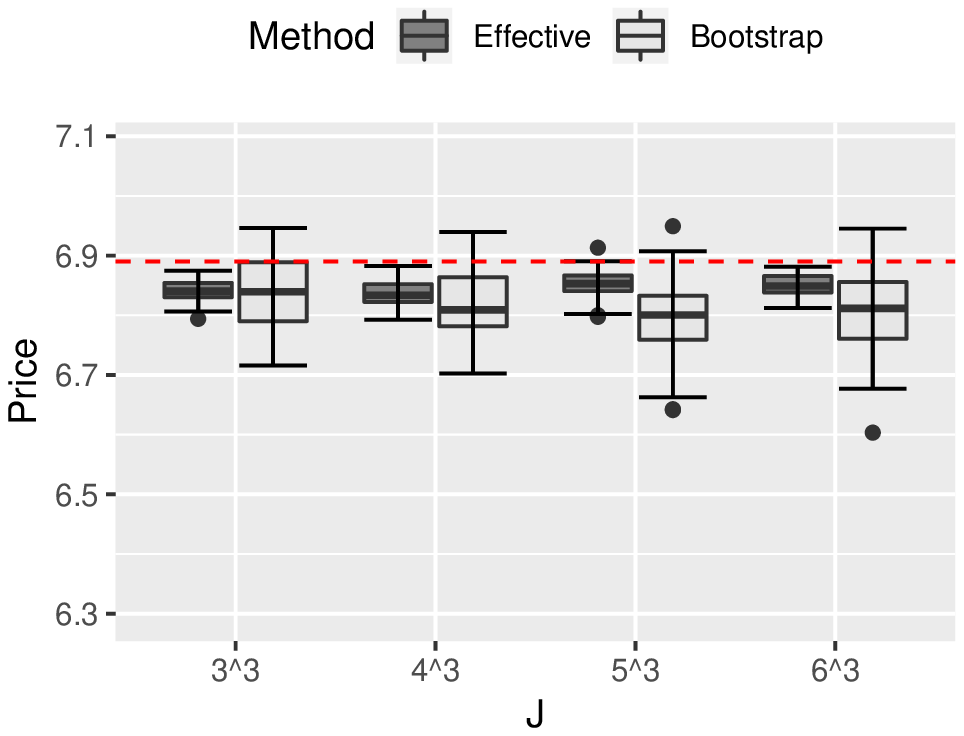}
		\caption{$N=5 \times 10^5$}
		\label{fig:PS3_ame_500}
	\end{subfigure}
	\caption{Price estimates of Asian options with early exercise (PS3), true price (6.89) indicated by the dashed line.}
	\label{fig:PS3_ame}
\end{figure}

We consider options with one year to maturity and use the same market and simulation parameters as in Section \ref{ssec:european_numerical_results}. 
Figures \ref{fig:PS2_ame} and \ref{fig:PS3_ame_100} present the prices obtained using two resampling methods: effective branching and bootstrap resampling (see Algorithm \ref{algo:bootstrap}).
In each case, the empirical distribution presented is estimated using 50 price estimates.
These results are compared to a ``ground truth price'' of 7.67 for PS2 and 6.89 for PS3, indicated by the red dashed line in the figures.
These prices were computed without resampling, but with a large number of simulations ($N=10^6$) and basis functions ($J=12^3$), to obtain a very precise result.

Table \ref{Tab:SADP_parameters} in the Appendix contains the SA/DP parameters used for all calculations. 
These parameters control the speed at which the gain step $\frac{\gamma}{k^\chi}$ goes to 0 as $k$ increases in the stochastic approximation algorithm. 
In our case, we do not want the gain steps to decrease too fast (that is, $\gamma$ should not be too small and $\chi$ should not be too large), as we want each of the $N$ paths to have some impact on the estimates $\bar{\alpha}^{J,N}_t$. 
In addition, since we use averaged coefficients $\bar{\alpha}^{J,N}_t$ to estimate the conditional expectations, we choose $\chi < 1$, as recommended in Section \ref{Pricing}.
Nonetheless, if $\gamma$ is too large, or if $\chi$ is too small, then we may not average out the noise fast enough and can run out of particles before getting to a good estimate for $\bar \alpha^{J,N}_t$. In \cite{Kouritzin16}, a stochastic search method was
used to find reasonable $\gamma$ and $\chi$.  However, the averaging method that we employ herein is far less sensitive to particular choice of $\gamma$ and $\chi$. Within the interval of values for which the gain step decreased at a satisfying rate, we ran the algorithm for different values of $\gamma$ and $\chi$ and chose those that minimized $\frac 1N \sum_{j=1}^N (Z^j_{\tau^{J,j}}-e^J(S^j_t,V^j_t)^\prime \bar{\alpha}^{J,N}_t)^2$ for $t \in {0,\ldots,T-1}$. 

Figures \ref{fig:PS2_ame} and \ref{fig:PS3_ame} clearly show that branching algorithms should be preferred over multinomial bootstrap when resampling is necessary to price American-type options. 
Indeed, the prices obtained with effective particle branching are in general closer to the true price than when bootstrap is used. 
For the second parameter set (PS2), prices obtained with resampling also show a clear convergence towards the true price as $J$ increases, while such a convergence is not observed when using bootstrap resampling.
The prices obtained with the third parameter set (PS3) are not as clear in terms of convergence; a larger number of simulations is needed to reduce the bias of the price estimate.

It has been shown theoretically in \cite{DeKoMi} that bootstrap resampling kills the process distribution of the simulated paths as the number of simulations gets very large.
Our results support this empirically; the price estimates obtained using bootstrap do not converge as clearly to the true price, and the variance of the estimator is in all cases significantly higher than with effective particle branching. 
For this reason, branching algorithms, and effective branching in particular, should be preferred when importance sampling is used for pricing options with early exercise.
Resampling algorithms based on particle branching are also typically faster than bootstrap algorithms, since less resampling occurs.

\section{Conclusions and Future Work} \label{Conclude}

Herein, we show that the simulation method presented by \cite{Kouritzin16} can be improved by resampling, especially when $\frac{4\nu}{\kappa^2}$ is close to 2. 
Since bootstrap resampling affects the process distribution of the particle system, the branching algorithms of \cite{BranchDraft}, also presented in Section \ref{BranAlg} in the context of option pricing, are a better alternative. 
In particular, effective particle branching offers more flexibility and allows the level of resampling to be a function of the state of the system.
We also highlight that the use of stochastic approximation algorithms within the dynamic programming framework for American option pricing is still valid when historical price paths are obtained via branching resampling algorithms.
Our use of the average coefficients $\bar{\alpha}^{J,N}$ in the pricing algorithm increases the robustness of the algorithm to the choice of parameters $\gamma$ and $\chi$. 
This is an important improvement over the algorithm used by \cite{Kouritzin16}, as it increases the stability of the pricing method and makes it easier to use, since less work is needed to find the optimal parameters. 

Future work will explore more advanced ways for the algorithms to ``learn'' the optimal resampling parameters.
Additional improvements linked to recycling simulated random variables in order to speed up the simulation algorithm will also be considered.

\appendix

\section{Additional Information on Numerical Results}

Table \ref{Tab:SADP_parameters} contains the stochastic approximation parameters used to obtain the numerical results of Section \ref{ssec:asian_numerical_results}.

\begin{table}[h!]
	\centering
	\small
	\caption{SA/DP parameters used for numerical examples.}
	\label{Tab:SADP_parameters}
	\begin{tabularx}{\textwidth}{*{5}{Y}}
		\toprule
		~ & \multicolumn{2}{c}{PS2} & \multicolumn{2}{c}{PS3} \\
		~ & Branching & Bootstrap & Branching & Bootstrap\\
		\cmidrule{1-5}
		&\multicolumn{4}{c}{$J=3^3$}\\
		\cmidrule{1-5}
		$\gamma$ & 3.0 & 2.0 & 3.0 & 2.0\\
		$\chi$ & 0.05 & 0.05 & 0.05 & 0.10\\
		\cmidrule{1-5}
		&\multicolumn{4}{c}{$J=4^3$}\\
		\cmidrule{1-5}
		$\gamma$ & 2.0 & 1.0 & 2.0 & 1.0\\
		$\chi$ & 0.10 & 0.10 & 0.10 & 0.10\\
		\cmidrule{1-5}
		&\multicolumn{4}{c}{$J=5^3$}\\
		\cmidrule{1-5}
		$\gamma$ & 1.0 & 0.5 & 1.0 & 0.5\\
		$\chi$ & 0.05 & 0.05 & 0.10 & 0.10\\
		\cmidrule{1-5}
		&\multicolumn{4}{c}{$J=6^3$}\\
		\cmidrule{1-5}
		$\gamma$ & 0.5 & 0.25 & 0.8 & 0.5\\
		$\chi$ & 0.02 & 0.02 & 0.10 & 0.10\\
		\bottomrule
	\end{tabularx}
\end{table}


%

\end{document}